\documentclass{article}

\usepackage{PRIMEarxiv}

\usepackage[utf8]{inputenc} 
\usepackage[T1]{fontenc}    
\usepackage{hyperref}       
\usepackage{url}            
\usepackage{booktabs}       
\usepackage{amsfonts}       
\usepackage{nicefrac}       
\usepackage{microtype}      
\usepackage{lipsum}
\usepackage{fancyhdr}       
\usepackage{graphicx}       
\graphicspath{{media/}}     
\usepackage[square,numbers]{natbib}
\usepackage{setspace}
\usepackage{indentfirst}
\usepackage{float}
\usepackage{caption}
\usepackage[belowskip=-10pt,aboveskip=5pt]{caption}
\usepackage{subcaption}
\usepackage{relsize}
\usepackage{mathrsfs,amsmath}
\usepackage{mathtools}
\usepackage{algorithm}
\usepackage{algpseudocode} 
\usepackage{color,soul}

\hypersetup{
	colorlinks=true,
	linkcolor=blue,
	citecolor=blue,
	urlcolor=blue,
	pdftitle={Multi-Level Bayesian Calibration of a Multi-Component Dynamic System Model},
	pdfkeywords={Information fusion, Bayesian statistics, Model calibration, Uncertainty quantification, Bayesian network, Machine learning},  
	pdfauthor={Berkcan Kapusuzoglu, Sankaran Mahadevan, Shunsaku Matsumoto, Yoshitomo Miyagi, Daigo Watanabe},
}

\title{Multi-Level Bayesian Calibration of a Multi-Component Dynamic System Model
\thanks{\textit{\underline{Citation}}: 
\textbf{Kapusuzoglu, B., Mahadevan, S., Matsumoto, S., Miyagi, Y., \& Watanabe, D. Multi-level Bayesian calibration of a multi-component dynamic system model. \textit{Journal of Computing and Information Science in Engineering} 23(1), 011006 (2023). DOI: \href{https://doi.org/10.1115/1.4055243}{10.1115/1.4055243}}
} 
}

\author{
  \textbf{Berkcan Kapusuzoglu}\thanks{Corresponding author: berkcan.kapusuzoglu@vanderbilt.edu}$^{\ ,1}$, \textbf{Sankaran Mahadevan}$^1$, \textbf{Shunsaku Matsumoto}$^2$, \\ \textbf{Yoshitomo Miyagi}$^3$, \textbf{Daigo Watanabe}$^2$ \\
  \vspace{0.2cm} \\
  $^1$Department of Civil and Environmental Engineering, Vanderbilt University, Nashville, TN 37235, USA \\
  $^2$Strength Research Department, Research and Innovation Center, Mitsubishi Heavy Industries, Ltd., \\ Nagasaki, 851-0392, Japan \\
  $^3$Strength Research Department, Research and Innovation Center, Mitsubishi Heavy Industries, Ltd., \\ Takasago, 676-8686, Japan
}

\begin{document}
\maketitle

\begin{abstract}
This paper proposes a multi-level Bayesian calibration approach that fuses information from heterogeneous sources and accounts for uncertainties in modeling and measurements for time-dependent multi-component systems. The developed methodology has two elements: quantifying the uncertainty at component and system levels, by fusing all available information, and corrected model prediction. A multi-level Bayesian calibration approach is developed to estimate component-level and system-level parameters using measurement data that are obtained at different time instances for different system components. Such heterogeneous data are consumed in a sequential manner, and an iterative strategy is developed to calibrate the parameters at the two levels. This calibration strategy is implemented for two scenarios: offline and online. The offline calibration uses data that is collected over all the time steps, whereas online calibration is performed in real-time as new measurements are obtained at each time step. Analysis models and observation data for the thermo-mechanical behavior of gas turbine engine rotor blades are used to analyze the effectiveness of the proposed approach.
\end{abstract}

\keywords{Information fusion \and Bayesian statistics \and Model calibration, Uncertainty quantification \and Bayesian network \and Physics-informed neural network (PINN)}

\section{Introduction}\label{sec:Intro}

In many engineering applications, physics-based computational models are used to analyze the system behavior under different operating conditions. The parameters of such models are not precisely known in some cases, and model predictions may not agree with observational data. Thus, model calibration needs to be performed to determine the values of unknown parameters, by comparing the model output to experimental data. Several model calibration techniques are available, such as  least squares, maximum likelihood estimation, and Bayesian estimation. The Bayesian approach provides a convenient framework for utilizing any prior knowledge about unknown parameters in model calibration ~\cite{vanderhorn2018bayesian,karve2020digital,viana2021survey}. 

The output of any model is affected by multiple sources of uncertainty; some of the sources are aleatory (due to natural variability) and some are epistemic (due to lack of knowledge). Uncertainty quantification (UQ) seeks to quantify the uncertainty in the model prediction arising from multiple, heterogeneous sources. UQ has two directions: the \textit{forward} problem and the \textit{inverse} problem. In the forward problem, model errors and the uncertainty related to the model inputs and parameters are propagated to compute the overall uncertainty in the output. In the inverse problem, model calibration (i.e., estimation of uncertain model parameters and errors) and system diagnosis (i.e., estimation of uncertain system states and errors) are performed using the available measurement data. The inverse problem is an essential part of uncertainty quantification and results in uncertainty reduction and in achieving the desired level of prediction confidence.

For coupled multi-physics systems with sparse data and many model parameters, the heterogeneous sources of uncertainty and their relationships need to be organized in a systematic manner to facilitate effective use of the available data. Bayesian networks (BNs) provide such a systematic approach, and represent the relationships between multiple models, their inputs and outputs, uncertainty sources, and experimental data~\cite{decarlo2016segmented,nannapaneni2016manufacturing}. BNs can facilitate analyses in both forward and inverse directions, thus supporting both the directions of UQ mentioned above. The uncertainty in the outputs can be estimated using forward propagation through the BN, by aggregating the uncertainty about various inputs, model parameters, and model errors. In the inverse problem, unmeasured inputs, system states, or model parameters can be estimated using Bayesian inference, given observations of some of the inputs and corresponding outputs. In general, BNs enable the inference of unmeasured quantities in a multi-physics or multi-component system model by using the observed data on measured quantities \cite{rebba2006model}.

In Bayesian calibration, prior and posterior beliefs about the calibration quantities are represented as probability density functions \cite{mahadevan2001bayesian}. Several studies have addressed the strategy for Bayesian calibration and the selection of priors and model discrepancy term in calibration \cite{kennedy2001bayesian,ling2014selection}. Recent studies have explored hierarchical approaches and Bayesian networks (BNs) \cite{sankararaman2015integration,li2016role, behmanesh2015hierarchical,nagel2015bayesian, nagel2016unified,jia2022hierarchical,sedehi2019probabilistic,SEDEHI2020106663,song2020accounting} for multilevel calibration. Nannanapeni et al. \cite{nannapaneni2016manufacturing} developed a hierarchical Bayesian network (HBN) approach for multi-model calibration for a time-independent system. For models sharing common calibration parameters, a probabilistic weighting approach was developed in \cite{sankararaman2015integration} for considering data from different experiments, based on validation results of the models used for calibration; this approach was extended in \cite{li2016role} to include the relevance of different types of measurements to the prediction quantities of interest. Mullins and Mahadevan \cite{mullins2016bayesian} developed a sequential approach to consider heterogeneous data from different sources in calibrating the model parameters.

The availability of data for model calibration is typically sparse in realistic systems, and data might be collected at different time instances for different system components. When multiple models are used to describe the system behavior (either multi-physics or multi-level), some unknown parameters are shared across multiple component models whereas some other unknown parameters are local to a given component. Also, some physical quantities might be measured for a single component and some other quantities for multiple components. In addition, the different measurements for different components could be asynchronous (i.e., available at different time instants, not all at the same instants). Asynchronous data is often encountered in the case of manual inspections, where different components or physical quantities may be measured at different times. The calibration of model parameters in situations where the models have complicated couplings between them and are based on different physics is not straightforward. Simultaneous Bayesian calibration of all the model parameters can be computationally prohibitive. Therefore, a segmented approach for Bayesian model calibration has been studied earlier for multi-physics problems \cite{decarlo2016segmented}, where two individual physics models with both individual and shared parameters were considered for a single component. However, in this case, the observation data on the outputs of each model were synchronous (i.e., available at the same time instants), they had one-to-one correspondence, since the output of one model was input to the other model, and they did not address issues related to time-dependent data. Later, DeCarlo et al. {\cite{decarlo2018quantifying}} focused on the resolution of the discrepancy models for coupled physics problems. Authors demonstrated their approach using time-dependent nature of the aerothermal data without considering a hierarchical setting.

The above review shows that previous studies have focused either on the calibration of multiple interacting physics models that contain both local and shared parameters for a single component, or multi-level calibration for a time-independent system with measurements of multiple output quantities. In these studies, either the system is static or the data is synchronous for the multiple measured quantities (with one-to-one correspondence). This motivates the development of a hierarchical Bayesian calibration methodology that incorporates several models that contain component-level and system-level parameters (in other words, local and shared parameters) and fuses heterogeneous data from multiple components at different time instants and spatial locations.

In this paper, we propose an approach that fuses information from multiple types of measurement data for Bayesian calibration of multi-physics and multi-component models of time-varying systems. Several practical scenarios are considered: (i) availability of measurement data on multiple output quantities of interest; (ii) some data measured only for a few components, and some data measured for all components; and (iii) different quantities measured at different time instants. Specific contributions are made in the following steps: (1) The methodology incorporates heterogeneous data scenarios listed from (i) to (iii) above. (2) An iterative calibration strategy is developed to estimate the component-level and system-level parameters, using a hierarchical BN and particle filter sampling approach. (3) The methodology is developed for both offline and online calibration. Offline calibration uses data that is collected over multiple time steps; whereas online calibration is performed in real-time whenever new measurements are obtained, thus continuously updating the model. The online strategy is also able to track changes in time-varying parameters. These contributions are valuable in constructing digital twins of multi-component systems. The proposed methods are demonstrated for the calibration of parameters for a thermo-mechanical model of gas turbine engine rotor blades.

The rest of this paper is organized as follows. A brief introduction to Bayesian networks and Bayesian calibration is given in Section \ref{Sec:backg}. The proposed methodology for multi-level Bayesian information fusion for time-dependent systems is presented in Section \ref{Sec:Methodology}. Section \ref{Sec:Numerical example} demonstrates the application of the proposed method for the calibration of thermo-mechanical model parameters (with spatio-temporal, multivariate output) for gas turbine engine rotor blades. Section \ref{Sec:Conclusion} provides concluding remarks, summarizes the contributions of this work, and identifies future research needs.

\section{Background}\label{Sec:backg}

\subsection{Bayesian Networks}\label{sec:BayesNetwork}
A Bayesian network (BN) is a directed acyclic graph (DAG) representation of a multivariate distribution, consisting of nodes and arcs, where nodes represent the random variables in the system and arcs (directed edges between nodes) are associated with conditional probabilities relating the nodes. In other words, a BN expresses the joint probability distribution of a set of variables through a set of conditional and marginal probabilities~\cite{pearl1988probabilistic}. Given random variables $X=\{X_1,X_2,...,X_n\}$, the joint probability of these variables is expressed through a BN as
\begin{equation}
    Pr(X) = \Pi_{i=1}^n Pr(X_i\lvert \Pi_{X_i})
\end{equation}
where $\Pi_{X_i}$ denotes the set of parent nodes of $X_i$ and $Pr(X_i\lvert \Pi_{X_i})$ represents the conditional probability distribution of $X_i$, given its parent nodes. If $X_i$ has no parent nodes (i.e.,  $Pr(X_i\lvert \Pi_{X_i}) = Pr(X_i)$ , then $X_i$ is a root node and is represented by a marginal distribution.

\subsection{Bayesian Model Calibration}\label{sec:bayesiancalibration}
Consider a physics model $G(\boldsymbol{\cdot})$ that maps input variables $\mathbf{X}$ and model parameters $\boldsymbol{\psi}$ to the model output $\mathbf{Y}_{m}$:
\begin{equation}
\mathbf{Y}_{m}(\mathbf{X}) = G\big(\mathbf{X};\ \boldsymbol{\psi}(\mathbf{X})\big)
\end{equation}
Let $n_{D}$ be the number of collected observation data $\mathbf{Y_{\mathrm{obs}}}$ corresponding to input variable settings $\mathbf{x}^{(1)}, ..., \mathbf{x}^{(n_{D})}$, where $\mathbf{x}^{(i)}$ is the input variable setting for the $i$th experiment. The difference between observations $\mathbf{Y}_{\rm obs}$ and the true response of the system $\mathbf{Y}_{\rm true}$ is attributed to measurement error $\epsilon_{\rm obs}$, which is often treated as a zero-mean Gaussian random variable with variance $\sigma_{\rm obs}^2$. The physics model prediction is inaccurate due to missing physics or other approximations. Thus, an additive model discrepancy term $\boldsymbol\delta(\mathbf{X})$ as a function of model inputs can be introduced to capture the difference between $\mathbf{Y}_{m}$ and the true response of the system $\mathbf{Y}_{\rm true}$~\cite{kennedy2001bayesian}:
\begin{align}
    &\mathbf{Y}_{\rm obs}(\mathbf{X}) = \mathbf{Y_{\mathrm{true}}}(\mathbf{X}) + \epsilon_{\rm obs}(\mathbf{X}), \label{eq:true_response} \\
    &\mathbf{Y}_{\rm true}(\mathbf{X}) = \mathbf{Y_{\mathrm{m}}}(\mathbf{X}) + \boldsymbol\delta(\mathbf{X}).
    \label{eq:modelerror}
\end{align}
Combining Eqs.~\eqref{eq:true_response} and~\eqref{eq:modelerror}, the overall prediction $\mathbf{Y}_{\rm pred}$ that accommodates various errors can be written as 
\begin{align}
    &\mathbf{Y}_{\rm true}(\mathbf{X}) = \mathbf{Y}_{\rm obs}(\mathbf{X}) - \epsilon_{\rm obs}(\mathbf{X}) = \mathbf{Y}_{m}(\mathbf{X}) + \boldsymbol\delta(\mathbf{X}), \nonumber\\
    &\mathbf{Y}_{\rm pred}(\mathbf{X}) = \mathbf{Y}_{m}(\mathbf{X}) + \boldsymbol\delta(\mathbf{X}) + \epsilon_{\rm obs}(\mathbf{X}).
    \label{eq:combined}
\end{align}
The unknown parameters of the physics model $\boldsymbol{\psi}$ and the parameters defining the model discrepancy term can be estimated using the experimental data, which contain measurement error $\epsilon_{obs}$, through Bayesian calibration. Thus the purpose of Bayesian model calibration is to use observation data $\mathbf{Y}_{\rm obs}$ to estimate the posterior distributions of unknown parameters of the physics model, the model discrepancy term, and the measurement error. The measurement error $\epsilon_{obs}$ is commonly represented as a zero-mean Gaussian random variable with an unknown variance $\sigma^2_{obs}$, i.e., $\epsilon_{obs}\sim N(0,\sigma^2_{obs})$. The model discrepancy $\delta$ can be formulated in various ways depending on the problem and can be a function of the input or any other parameter. Thus $\boldsymbol\delta(\mathbf{X})$ may be modeled as a constant bias term, or a random variable with either fixed or input-dependent mean and variance, or a random process~\cite{ling2014selection}. When the model parameters $\boldsymbol{\psi}$ are uncertain, the calibration quantities are $\boldsymbol{\Theta}=[\boldsymbol{\psi},\boldsymbol{\theta},\sigma_{obs}]$, where $\boldsymbol{\theta}$ denotes the parameters of $\delta$. Based on Bayes' theorem, the joint distribution of the calibration quantities is given by
\begin{equation}
    f(\boldsymbol{\Theta}|\mathbf{Y}_{\rm obs})=\frac{f(\mathbf{Y}_{\rm obs}|\boldsymbol{\Theta})f(\boldsymbol{\Theta})}{\int f(\mathbf{Y}_{\rm obs}|\boldsymbol{\Theta})f(\boldsymbol{\Theta})d\boldsymbol{\Theta}}
\end{equation}\label{eq:postlikeli}
where $\mathbf{Y}_{\rm obs}$ is the observation data, $f(\mathbf{Y}_{\rm obs}|\boldsymbol{\Theta})$, $f(\boldsymbol{\Theta})$, and $f(\boldsymbol{\Theta}|\mathbf{Y}_{\rm obs})$ are respectively the likelihood function, the joint prior PDF (probability density function) of $\boldsymbol{\Theta}$, and the joint posterior PDF of $\boldsymbol{\Theta}$.

Markov Chain Monte Carlo (MCMC) sampling~\cite{hastings1970monte} is commonly used to estimate the posterior distributions of the calibration quantities. The method requires a large number of samples to construct the posterior, and can become computationally unaffordable in an online sequential Bayesian estimation context \cite{Moral2006} (i.e., model updating at multiple time instants). Thus, a particle filter sampling algorithm \cite{fearnhead2004particle}, which is commonly used for Bayesian update of system states for dynamic systems, is adopted in this paper to calibrate the model parameters. Since the high-fidelity physics model is very expensive, an inexpensive surrogate model, which makes Bayesian calibration affordable, is built and used.

\section{Proposed Methodology}\label{Sec:Methodology}

In many practical engineering systems, accurately estimating the remaining useful life and real-time system response is a challenge due to the complexity of physics and uncertainty in service environments. A large number of computationally expensive high-fidelity simulation model runs (e.g., FEA, CFD) with different samples of model inputs and parameters need to be performed in order to quantify the effect of multiple uncertainty sources on the QoI. This is often unaffordable, therefore surrogate models need to be constructed to replace these computationally expensive simulation models. Further, numerical computational models such as FEA and CFD provide multiple output quantities (e.g., stresses and deformations along several degrees of freedom) at a large number of locations and time instants. When the output of the physics-based model is high-dimensional and spatio-temporal, the direct use of all the available data simultaneously for Bayesian calibration is not practical.

A hierarchical Bayesian network (HBN) approach is used to integrate expert knowledge and measurements from multiple domains at different levels of the network. HBNs offer several advantages for modeling, analysis, and visualization of complex processes. The BNs corresponding to a simpler component (coupon) can be constructed separately and composed to form an HBN that represents a more complex system. HBN allows the integration of expert knowledge from multiple domains in modeling the complex system. HBNs also enable scalability in modeling such a complex system. As the system becomes more complex, the number of simpler components and more complex parts increase. HBNs allow modeling of all these different levels of processes in a systematic way.

The proposed methodology aims to develop an efficient Bayesian calibration approach in time-dependent systems through multi-source information fusion. Multiple types of measurement data from multiple components at different time instants are considered. Two calibration strategies are investigated: offline and online. The information fusion at different levels is presented for both strategies and the differences are highlighted. 

\subsection{Bayesian Calibration with Multiple Types of Data}\label{sec:BC_multi}
The measurement data $\mathcal{D}_{obs}$, for Bayesian calibration can be available for different components at different time instants and spatial locations in a multi-component system. Some of the model parameters can be local $\boldsymbol{\psi}_{local}$ (i.e., a unique value for each component) and others can be global $\boldsymbol{\psi}_{global}$ (i.e., a common value across components). Simultaneous calibration of all unknown parameters $\boldsymbol{\Theta}$ can be computationally prohibitive with multi-component, asynchronous data without the use of a modular Bayesian calibration approach. Therefore, an efficient methodology is proposed here to perform multi-level Bayesian calibration by fusing different data types from multiple components.

\begin{figure}[t]
\centering
    \includegraphics[width=0.45\textwidth]{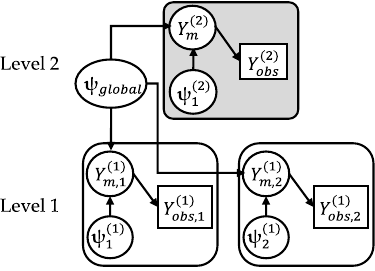}
  \caption{A simple hierarchical Bayesian network (\emph{static})}
  \label{fig:HBN_}
\end{figure}

\begin{figure*}[t]
\centering
    \includegraphics[width=0.8\textwidth]{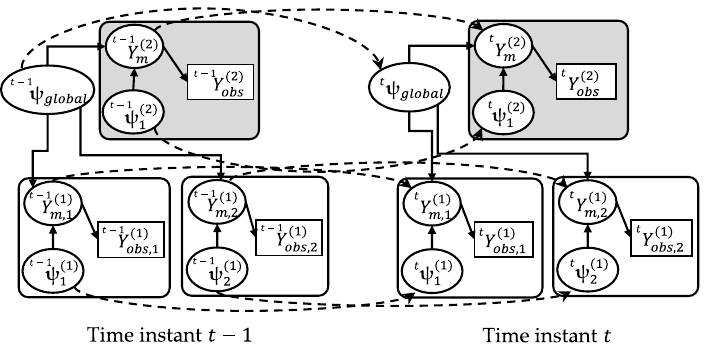}
  \caption{A simple hierarchical \emph{dynamic} Bayesian network}
  \label{fig:DBN_}
\end{figure*}

The proposed multi-level calibration approach leverages a hierarchical Bayesian network (HBN) where some nodes in a BN may represent another lower-level BN as shown in Fig. \ref{fig:HBN_} (with superscripts denoting the level number). Some of the nodes in the lower-level BN can also represent further lower-level BNs, thus any number of levels are possible. Note that Figs. \ref{fig:HBN_} and \ref{fig:DBN_} do not include other input variables and intermediate quantities that would normally be part of the BN for a realistic system, in order to avoid crowding of the figure. Only the model parameters, model outputs and observations are included in the figures since those are directly considered in the discussion below. The inverse problem of Bayesian calibration is achieved by passing the information from the data nodes (solid squares) to the calibration quantities. The observation data in each lower-level BN (i.e., $Y_{obs,1}^{(1)}$ and $Y_{obs,2}^{(1)}$, corresponding to model predictions $Y_{m,1}^{(1)}$ and $Y_{m,2}^{(1)}$, respectively) are used to estimate the posterior distributions of the nodes in the lower-level BN (e.g., $\psi_1^{(1)}$, and $\psi_2^{(1)}$) and the shared parameters $\boldsymbol{\psi}_{global}$. During calibration of the nodes using the data $Y_{obs}^{(2)}$ in the higher-level BNs (corresponding to model prediction $Y_{m}^{(2)}$), the posterior distributions from the lower-level calibration are used as prior distributions to re-calibrate those parameters that also go into higher-level BNs.

The approach described in the preceding paragraph is directly applicable to a static (time-independent) Bayesian network. Extension to a dynamic Bayesian network (DBN) with time-dependent system states and parameters is given in Fig. \ref{fig:DBN_}. The subscripts $t-1$ and $t$ denote the time instants, and the nodes in the DBN are connected by arrows that represent conditional probability distributions or deterministic functional relations. In addition to the dynamic nodes for which the states change over time, there can also be static nodes that are included at all time instances. The problem considered here is simpler, only model parameter estimation and no updating of uncertain time-dependent system states; thus we only consider the estimation of static parameters (i.e., they do not change over time). Note that Fig. \ref{fig:DBN_} does not include all the time transitions between the quantities at time $t-1$ and time $t$, in order to avoid crowding of the figure.

Denoting the vector of all calibration parameters as $\boldsymbol{\psi}$, let the vectors $\boldsymbol{\psi}^{(1)}$ and $\boldsymbol{\psi}^{(2)}$ denote the calibration parameters at Levels 1 and 2, and let $\mathcal{D}_{BN_1}=\{Y_{obs,1}^{(1)},Y_{obs,2}^{(1)}\}$ represent the data available at Level 1 (corresponding to model predictions $Y_{m,1}^{(1)}$ and $Y_{m,2}^{(1)}$, respectively), and let $\mathcal{D}_{BN_2}=Y_{obs}^{(2)}$ represent the data available at Level 2 (corresponding to model prediction $Y_{m}^{(2)}$). We consider two cases: (1) the available data $\mathcal{D}_{BN_1}$ and $\mathcal{D}_{BN_2}$ are independent, or (2) there is a one-to-one correspondence between them. Assuming $\mathcal{D}_{BN_1}$ and $\mathcal{D}_{BN_2}$ are independent, the posterior distributions $\Pi(\boldsymbol{\psi}^{(1)},\boldsymbol{\psi}^{(2)}\lvert \mathcal{D}_{BN_1},\mathcal{D}_{BN_2})$ can be obtained as
{\footnotesize
\begin{align}\label{eq:firstcase}
    \Pi&(\boldsymbol{\psi}^{(1)},\boldsymbol{\psi}^{(2)}\lvert \mathcal{D}_{BN_1},\mathcal{D}_{BN_2}) \nonumber\\
    &\propto L(\mathcal{D}_{BN_1},\mathcal{D}_{BN_2}\lvert \boldsymbol{\psi}^{(1)},\boldsymbol{\psi}^{(2)}) \Pi(\boldsymbol{\psi}^{(1)},\boldsymbol{\psi}^{(2)}) \nonumber\\
    &\propto L(\mathcal{D}_{BN_1}\lvert \boldsymbol{\psi}^{(1)}) L(\mathcal{D}_{BN_2}\lvert \mathcal{D}_{BN_1},\boldsymbol{\psi}^{(1)},\boldsymbol{\psi}^{(2)})\Pi(\boldsymbol{\psi}^{(1)})\Pi(\boldsymbol{\psi}^{(2)})\nonumber\\
    &\propto L(\mathcal{D}_{BN_1}\lvert \boldsymbol{\psi}^{(1)})\Pi(\boldsymbol{\psi}^{(1)}) L(\mathcal{D}_{BN_2}\lvert \boldsymbol{\psi}^{(1)},\boldsymbol{\psi}^{(2)})\Pi(\boldsymbol{\psi}^{(2)})
\end{align}
}
\noindent where the first two terms in the last expression denote the posterior distributions of $\boldsymbol{\psi}^{(1)}$, $\Pi(\boldsymbol{\psi}^{(1)}\lvert \mathcal{D}_{BN_1})$, using the lower-level data $\mathcal{D}_{BN_1}$. The posterior distributions of $\boldsymbol{\psi}^{(1)}$ are then used as prior distributions to obtain the posterior distributions of $\boldsymbol{\psi}^{(1)}$ and $\boldsymbol{\psi}{(2)}$ using the higher-level data $\mathcal{D}_{BN_2}$. Since $\mathcal{D}_{BN_1}$ and $\mathcal{D}_{BN_2}$ are assumed to be independent, the likelihood term $L(\mathcal{D}_{BN_2}\lvert \mathcal{D}_{BN_1},\boldsymbol{\psi}^{(1)},\boldsymbol{\psi}^{(2)})$ simplifies to $L(\mathcal{D}_{BN_2}\lvert \boldsymbol{\psi}^{(1)},\boldsymbol{\psi}^{(2)})$. In other words, the posterior distributions obtained in the lower-level BNs using data at that level are propagated as priors to the higher-level BN for re-calibration using the higher-level data $\mathcal{D}_{BN_2}$. 

In the second case, the data $\mathcal{D}_{BN_1}$ and $\mathcal{D}_{BN_2}$ are assumed to have a one-to-one correspondence (i.e., there exists a corresponding $\mathcal{D}_{BN_1}$ for a given $\mathcal{D}_{BN_2}$); then the posterior distribution is given as
{\footnotesize
\begin{align}
    \Pi&(\boldsymbol{\psi}^{(1)},\boldsymbol{\psi}^{(2)}\lvert \mathcal{D}_{BN_1},\mathcal{D}_{BN_2}) \nonumber\\
    &\propto L(\mathcal{D}_{BN_1},\mathcal{D}_{BN_2}\lvert \boldsymbol{\psi}^{(1)},\boldsymbol{\psi}^{(2)}) \Pi(\boldsymbol{\psi}^{(1)},\boldsymbol{\psi}^{(2)}) \nonumber\\
    &\propto L(\mathcal{D}_{BN_1}\lvert \boldsymbol{\psi}^{(1)}) L(\mathcal{D}_{BN_2}\lvert \mathcal{D}_{BN_1},\boldsymbol{\psi}^{(1)},\boldsymbol{\psi}^{(2)})\Pi(\boldsymbol{\psi}^{(1)})\Pi(\boldsymbol{\psi}^{(2)})\nonumber\\
    &\propto L(\mathcal{D}_{BN_1}\lvert \boldsymbol{\psi}^{(1)})\Pi(\boldsymbol{\psi}^{(1)}) L(\mathcal{D}_{BN_2}\lvert \mathcal{D}_{BN_1},\boldsymbol{\psi}^{(2)})\Pi(\boldsymbol{\psi}^{(2)})
\end{align}
}
\noindent where the likelihood term $L(\mathcal{D}_{BN_2}\lvert \mathcal{D}_{BN_1},\boldsymbol{\psi}^{(1)},\boldsymbol{\psi}^{(2)})$ simplifies to $L(\mathcal{D}_{BN_2}\lvert \mathcal{D}_{BN_1},\boldsymbol{\psi}^{(2)})$ since $\boldsymbol{\psi}^{(2)}$ is independent of $\boldsymbol{\psi}^{(1)}$ when $\mathcal{D}_{BN_1}$ is known. The first two terms provide the posterior distributions of $\boldsymbol{\psi}^{(1)}$ using $\mathcal{D}_{BN_1}$ similar to the previous case, and the last two terms provide the posterior distributions of $\boldsymbol{\psi}^{(2)}$ using $\mathcal{D}_{BN_1}$ and $\mathcal{D}_{BN_2}$. Thus, the final posterior distributions of $\boldsymbol{\psi}^{(1)}$, $\boldsymbol{\psi}^{(2)}$ can be obtained in one shot separately, whereas a two-step approach is used in the earlier case described in Eq. \eqref{eq:firstcase} to obtain the posterior of $\boldsymbol{\psi}^{(1)}$: Step 1: Calibrate using only $\mathcal{D}_{BN_1}$; Step 2: Use the posterior from step 1 to re-calibrate together with $\boldsymbol{\psi}^{(2)}$ using $\mathcal{D}_{BN_2}$. The above approaches can be extended to calibrate hierarchical DBNs with multiple levels. This paper pursues the two-step approach to estimate not only the component-level and system-level model parameters but also the model errors in multi-level transient systems.

Multi-level hierarchy can be in terms of either the models, or measurement data, or both. In the case of models, the output of one model (e.g., FEM) in the lower-level may go to a second model (e.g., creep model) in the higher-level. In terms of data, some quantities may be measured for multiple components (e.g., deformation), whereas some other quantities may be measured only for a few specific component (e.g., creep damage); these different data types may be handled in different levels of the multi-level calibration approach. The numerical example in Section \ref{Sec:Numerical example} features both types of hierarchy, and uses the different data types for multi-level calibration of the model parameters.

\subsection{Offline and Online Multi-Level Calibration}\label{sec:off_on}
The multi-level Bayesian calibration approach developed above can be pursued in two ways: Offline and Online (see Algorithm~\ref{algorithm}). In the former strategy, the calibration of all parameters is performed at once using all the available data, with data on different components that are collected at different time steps. In the latter strategy, the calibration is performed in real-time whenever measurements are obtained. In the offline strategy, $\boldsymbol{\psi}_{global}$ are fixed at their priors and $\boldsymbol{\psi}_{local}$ are calibrated for each system component using measurements at each time step and the posterior distributions from this calibration are then used as prior distributions to calibrate the global parameters while fixing the non-global parameters at their posterior estimates from the previous iteration step. Whereas in the online strategy, the same procedure is performed only at the time step when measurements are taken. The posteriors obtained from this data type are propagated in time and used as priors to further update the parameter estimates using subsequent measurements. As new measurements are taken, model parameter distributions are updated; thus the data are used sequentially as they become available.

The calibration strategy as described in Algorithm~\ref{algorithm} consists of the following general steps:
\begin{enumerate}
\item First assume priors for the parameters to be calibrated (both local and global parameters);
\item Next, use data collected from multiple components to calibrate both $\boldsymbol{\psi}_{local}$ and $\boldsymbol{\psi}_{global}$ in an iterative manner until convergence;
\begin{enumerate}
\item First fix $\boldsymbol{\psi}_{global}$ at their priors (no updating) and $\boldsymbol{\psi}_{local}$ are updated for each sample;
\item Then, $\boldsymbol{\psi}_{local}$ are fixed at their posterior distributions (no updating) and $\boldsymbol{\psi}_{global}$ are updated;
\item Steps 2(a) and 2(b) are then repeated until convergence, i.e., until the change in the mean and variance of the posterior distribution of $\boldsymbol{\psi}_{global}$ is smaller than a set threshold. (This is similar to the well-known expectation maximization (EM) algorithm~\cite{dempster1977maximum});
\end{enumerate}
\item The posterior distributions obtained in step 2(c) are used as priors for further re-calibration of calibration parameters using the single-component data.
\end{enumerate}
\begin{algorithm}[t]
  \caption{\footnotesize Multi-level Bayesian calibration pseudo code: \\ (A) Offline calibration, and (B) Online calibration}\label{algorithm}
  \footnotesize
  \textbf{Given}: $\mathcal{D}_{obs}$\\
  \textbf{Output}: Posterior estimates of $\boldsymbol{\psi}$\
  \begin{algorithmic}[1]
  
    \Procedure{(a) Offline Calibration}{}
    \For{$i= 1\ \text{to}\ n_{time}$}\Comment{Multi-component data}
        \While{(Change in posteriors of $\boldsymbol{\psi}_{global}$ $>$ \emph{threshold})}
            \State{Fix $\boldsymbol{\psi}_{global}$ at their priors}
            \State{Calibrate $\boldsymbol{\psi}_{local}$}
            \State{Fix $\boldsymbol{\psi}_{local}$ at their posterior and update $\boldsymbol{\psi}_{global}$}
        \EndWhile
    \EndFor
    \For{$i= 1\ \text{to}\ n_{time}$}\Comment{Single-component data}
        \State{Re-calibrate $\boldsymbol{\psi}_{local}$ and $\boldsymbol{\psi}_{global}$}
    \EndFor
    \EndProcedure
    \\
    \Procedure{(b) Online Calibration}{}
    \For{$i= 1\ \text{to}\ n_{time}$}
        \While{(Change in posteriors of $\boldsymbol{\psi}_{global}$ $>$ \emph{threshold})}\Comment{Multi-component data}
            \State{Fix $\boldsymbol{\psi}_{global}$ at their priors}
            \State{Calibrate $\boldsymbol{\psi}_{local}$}
            \State{Fix $\boldsymbol{\psi}_{local}$ at their posterior and update $\boldsymbol{\psi}_{global}$}
        \EndWhile
        \State{Re-calibrate $\boldsymbol{\psi}_{local}$ and $\boldsymbol{\psi}_{global}$}\Comment{Single-component data}
    \EndFor
    \EndProcedure
  \end{algorithmic}
\end{algorithm}
The main difference between the offline and online strategies is that the offline strategy has access to the complete measurement history. This allows the use of all available multi-component data for calibration (lines 2-8 in Algorithm~\ref{algorithm}). Whereas, in the online strategy the multi-component data is fed into the calibration strategy in real-time (lines 15-20 in Algorithm~\ref{algorithm}). This can lead to a different convergence behavior between the offline and online strategies since the result of iterative calibration procedure is dependent on the data that is used.

\subsection{Surrogate Modeling}\label{sec:ML_model}
The calibration strategy developed above requires multiple evaluations of expensive physics models; thus surrogate modeling is needed to reduce the computational effort. The inexpensive surrogate model can be constructed using a much smaller number of physics model runs. Polynomial chaos expansion (PCE) \cite{xiu2002wiener} and Gaussian process (GP) regression \cite{Rasmussen2004} models are commonly used in the UQ literature. In this paper, we explore both these techniques as well as other types of surrogate modeling techniques such as Extra-Trees regressor~\cite{geurts2006extremely}, Light Gradient Boosting Machine (lightGBM) \cite{NIPS2017_6449f44a}, Extreme Gradient Boosting (XGBoost) \cite{xgboost}, Categorical Boosting (CatBoost) \cite{dorogush2018catboost}, and random forest (RF) \cite{hastie2009elements}, and select the best model among them by comparing their accuracy (characterized by root mean square error, RMSE) on validation data separate from training data. 

Further, a large multi-component system may have a very high-dimensional output. Singular value decomposition (SVD) could be used to map the high-dimensional output to a low-dimensional feature space and the surrogate model could be built in the feature space. However, the basic SVD method becomes infeasible for very high-dimensional outputs due to the memory demands in storing and inverting very large matrices. Therefore, we employed the randomized singular value decomposition (rSVD) \cite{halko2011finding} approach to identify the important features in the high-dimensional output space and give a lower-dimensional representation of the original QoIs. These important features are then used as the outputs of the surrogate model; thus the surrogate model is constructed in the low dimensional space. Using reconstruction (i.e., inversion of rSVD), the prediction of the trained surrogate model can be translated to the QoIs in the original space~\cite{kapusuzoglu2022dimension}.

In this paper, we define the initial physics model runs using a space-filling design of experiments (DOE) method called maximin Latin hypercube sampling (LHS) \cite{morris1995exploratory} to generate initial training points that explore the input space uniformly. The accuracy of the initial surrogate model that is built with these initial sets of inputs and the corresponding important features is evaluated using cross-validation. Additional training runs of the physics model are selected using a sequential adaptive sampling strategy, which considers both the physics of the system and the coverage of the input space, to populate new samples in regions of \emph{high interest} and the surrogate model is improved until it reaches the desired accuracy; thus optimizing resource allocation~\cite{kapusuzoglu2022dimension}.

\subsection{Summary of Methodology}\label{sec:summary}
The proposed methodology consists of the following elements: (1) Collection of physics simulation data for training the surrogate model; (2) Dimension reduction of the high-dimensional output; (3) Surrogate model construction in the low-dimensional space; (4) Collection of measurement data on the QoIs; (5) Information fusion and iterative calibration of the unknown local and global model parameters; and (6) Validation of the prediction using the calibrated model. In the next section, we demonstrate the effectiveness of the proposed methodology for a high-dimensional problem with thermo-mechanical analysis and spatio-temporal output.

\section{Numerical Example}\label{Sec:Numerical example}
\subsection{Introduction to the Problem}
A simplified gas turbine engine blade model is studied in this section to demonstrate the proposed surrogate modeling and calibration techniques for high-dimensional spatio-temporal output. The output quantities of interest (QOIs) of the FEM runs are creep equivalent strain (CEEQ), creep damage (Dc), von Mises stress, displacement in x, y, and z directions; these are available at a large number of spatial and temporal points. The four-dimensional input for the surrogate model consists of turbine blade coating thickness (TBC thickness), turbine output rate (T1T rate), transient and constant creep rate material properties, $\alpha$ and $\gamma$, respectively. The ranges of the inputs are: TBC thickness [2/3,4/3], T1T rate [80,100], $\alpha$ and $\gamma$ [-3,3]. The boundary conditions of the FEM model are assumed to be known with certainty. The total number of spatial locations of the FEM model is over 27,000 for each output QoI and the total number of time steps is 54. Thus, for each QoI, a single FEM simulation gives a vector of dimension over 1,458,000.

\subsection{Surrogate Modeling}
The surrogate model needs to be able to jointly predict all the QoIs while taking the correlation between them into account and improving computational efficiency~\cite{kapusuzoglu2022dimension}. Among the surrogate modeling methods mentioned in Section \ref{sec:ML_model} Extra-Trees regressor~\cite{geurts2006extremely} was found to give the best results for this problem, with an average cross-validation (CV) RMSE value of 0.10513, which is 15\% smaller than the average CV RMSE of the second best surrogate model. The main idea behind Extra-Trees is to randomly generate a number of different trees with randomly chosen features. This randomization reduces the variance of the model. The four inputs to the Extra-Trees surrogate model are TBC thickness, T1T rate, $\alpha$, and $\gamma$, and the outputs of the surrogate model are the 20 important features. The number of important features is chosen based on the percentage of variance explained by the top few singular vector / singular value pairs and reconstruction accuracy. The percentage of variance explained by the top 20 features is 99\% and the root mean square error (RMSE) for the reconstruction accuracy is approximately 0.0006, which is less than 1\% of the average magnitude of the predicted quantities (0.01). The RMSE value does not change significantly by using more important features to approximate the original matrix. 

\begin{figure*}[h]
\centering
    \includegraphics[width=4.2in]{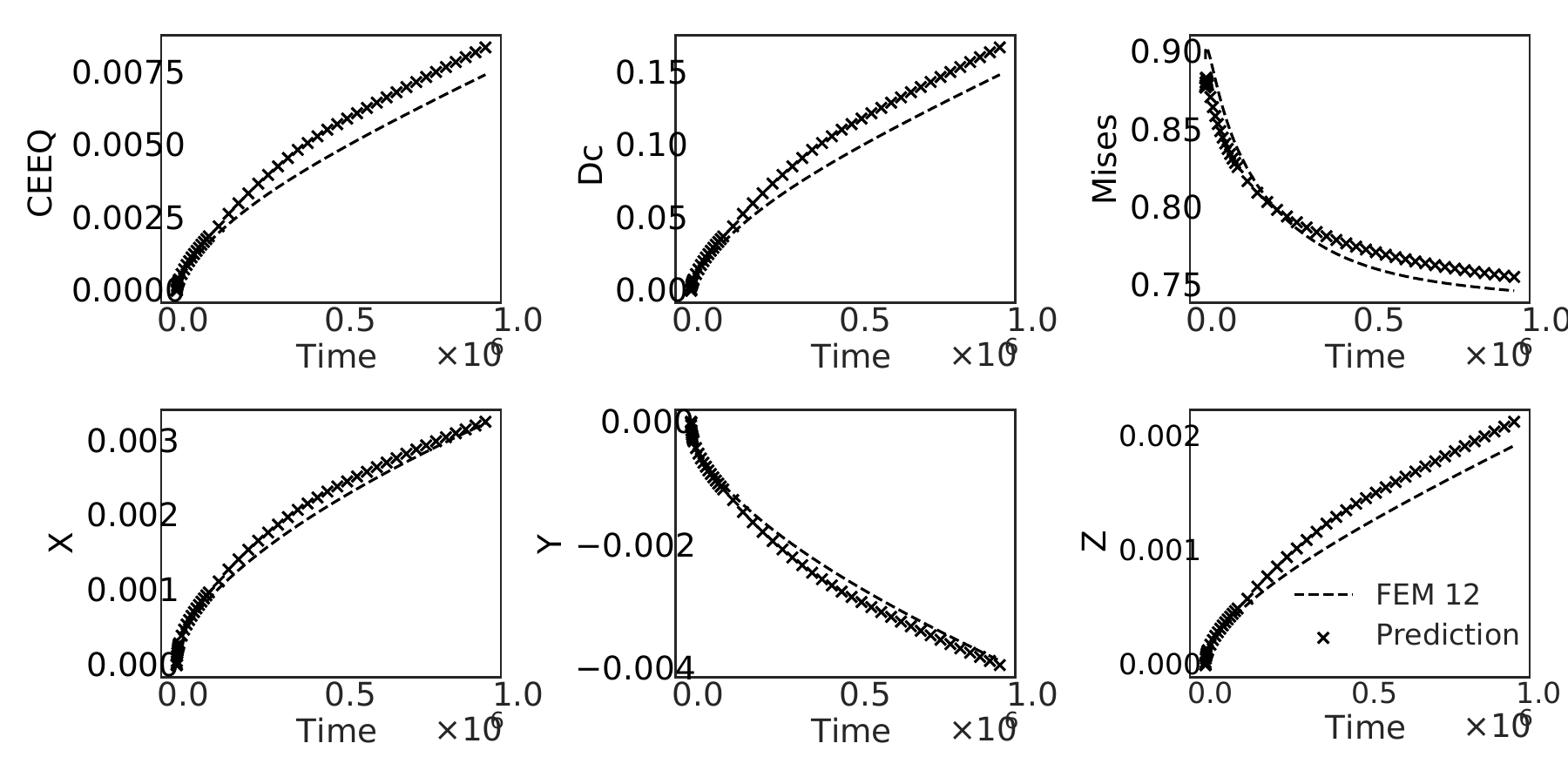}
  \caption{Surrogate model prediction compared against FEM 12 at the important node (id 5057)}
  \label{fig:pred_vs_fem37}
\end{figure*}

The $R^2$ value of the predictions on the test set in the low-dimensional space is 0.97. The surrogate model prediction accuracy is evaluated at an important node (id 5057; based on expert opinion) as shown in Fig.~\ref{fig:pred_vs_fem37}. The surrogate model predictions are observed to be in good agreement with the FEM results (similar trends in some cases if not actual values), which demonstrates the effectiveness of the surrogate model.
\subsection{Bayesian Parameter Calibration}
We use the particle filter algorithm to compute the posterior distributions of $\boldsymbol{\psi}=$ [TBC thickness, T1T rate, $\alpha$, $\gamma$]. A uniform distribution is chosen as the prior distribution for each parameter of $\boldsymbol{\psi}$. In sequential updating using the PF method, for each proposed $\boldsymbol{\psi}$, the important features are calculated using the corresponding surrogate model that is trained with 66 FEM runs. The physical quantities in the original space are subsequently calculated by performing an inverse projection of the important features onto the original space. The likelihood function $f(\mathbf{Y}_{\rm obs}|\boldsymbol{\Theta})$ is calculated while considering a stochastic observed measurement error $\epsilon_{obs}$ that is represented as a zero-mean Gaussian random variable with an unknown variance $\sigma^2_{obs}$, i.e., $\epsilon_{obs}\sim N(0,\sigma^2_{obs})$.

\begin{figure*}[h]
\centering
    \includegraphics[width=0.6\textwidth]{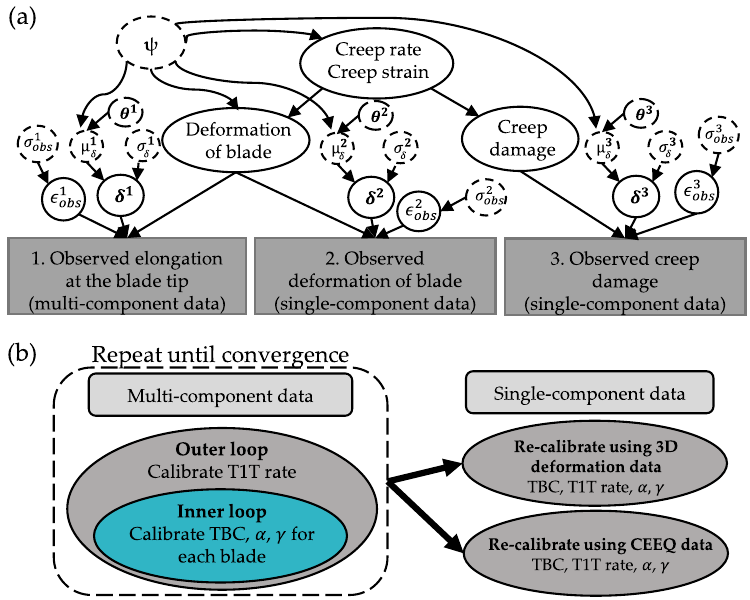}
  \caption{(a) Multi-level Bayesian network and (b) Calibration in multi-component system}
  \label{fig:steps_pf_multilevel}
\end{figure*}

There are over 27,000 spatial locations for each QoI (i.e., FEM nodes) and there are a total of 54 time steps. However, the measurement data is significantly sparse. We perform multi-level Bayesian calibration using different types of measurement data and multiple models (i.e., a coupled CFD-FEM model and a creep model) as described in Fig.~\ref{fig:steps_pf_multilevel} to make use of all the data. We have single-blade (single-component) data and multi-blade (multi-component) data. Elongation at the blade tip is available for 100 blades, and x, y, z displacement data is only available for a single blade (i.e., blade ID 01). Both elongation at the blade tip and x, y, z displacement data are collected at the same three time instants (see Table \ref{table:expdata}). For example, the elongation data at the blade tip (i.e., Node ID 31933) is collected for 100 blades and only at three time instants. The 3D deformation data is obtained at the same three time instants at 6448 different spatial locations for blade ID 01. We have CEEQ data only for three blades. CEEQ measurements are obtained by destructive testing, which is different from collecting displacement data with a laser scanner (non-destructive). After the destructive testing of a blade, another blade will be installed as replacement. Thus, CEEQ data for a single blade is only available at one time instant and at a single spatial location (i.e., Node ID 5057) as shown in Table \ref{table:expdata}.

The measurements were only collected at 3 time steps because of the assumptions made based on a realistic scenario. Turbine blades under high temperature are usually inspected during planned outage and replaced if the remaining useful life is short. The measurement data such as elongation and displacement was only available during inspection due to technical limitations (measurement under high temperature), thus measurement data was collected at sparse time steps in its lifetime.
\begin{table}[h]
    \caption{Experimental data}
    \label{table:expdata}
\centering{%
    \resizebox{0.485\textwidth}{!}{%
        \begin{tabular}{@{\extracolsep{\fill}}l*{4}c@{\extracolsep{\fill}}}
        \hline\noalign{\smallskip}
            & \multicolumn{3}{c}{Time instants} \\
            \cline{2-4}
            Data type & $t_1$ & $t_2$ & $t_3$ \\
            \noalign{\smallskip}\hline\noalign{\smallskip}
            Elongation at the blade tip & blade ID 01-100 & blade ID 01-100 & blade ID 01-100  \\
            XYZ deformation & blade ID 01 & blade ID 01 & blade ID 01  \\
            Creep equivalent strain (CEEQ) & blade ID 28 & blade ID 82 & blade ID 99 \\
        \noalign{\smallskip}\hline
        \end{tabular}
    }%
}%
\end{table}

The multi-level Bayesian calibration approach is implemented along the following steps: 
\begin{enumerate}
    \item Assume priors for the model parameters, where three of them $\boldsymbol{\psi}_{local} =$ \{TBC thickness, $\alpha$, $\gamma$\} are local parameters, i.e., unique to each blade, and $\boldsymbol{\psi}_{global}=$ \{T1T\} is a global parameter across all blades. In addition, assume priors for the standard deviation of measurement error $\sigma_{obs}$ and parameters $\boldsymbol{\theta}$ of input-output dependent model discrepancy term, $\delta$.
    \item Use multi-blade data, i.e., tip elongation of 100 blades, to calibrate $\boldsymbol{\psi}_{local}$, $\boldsymbol{\psi}_{global}$, $\sigma^1_{obs}$, and $\boldsymbol{\theta}^1$ in an iterative manner as explained in Section \ref{sec:off_on} and shown in Fig.~\ref{fig:steps_pf_multilevel}(b).
    \item Use the posterior distributions obtained in the above step as priors to re-calibrate $\boldsymbol{\psi}$ and calibrate $\sigma^2_{obs}$, and $\boldsymbol{\theta}^2$ using the single-blade 3D deformation data of blade ID 01 shown in Fig.~\ref{fig:steps_pf_multilevel}(b).
    \item Use the posterior distributions of $\boldsymbol{\psi}_{local}$ and $\boldsymbol{\psi}_{global}$ obtained from the second step (using the multi-blade data) as priors for re-calibrating $\boldsymbol{\psi}$ and calibrating $\sigma^3_{obs}$, and $\boldsymbol{\theta}^3$ using CEEQ measurements of blade IDs 28, 82, and 99 at three unique time instances (see Table \ref{table:expdata}).
\end{enumerate}

\begin{figure}[t]
\centering
    \includegraphics[width=2.85in]{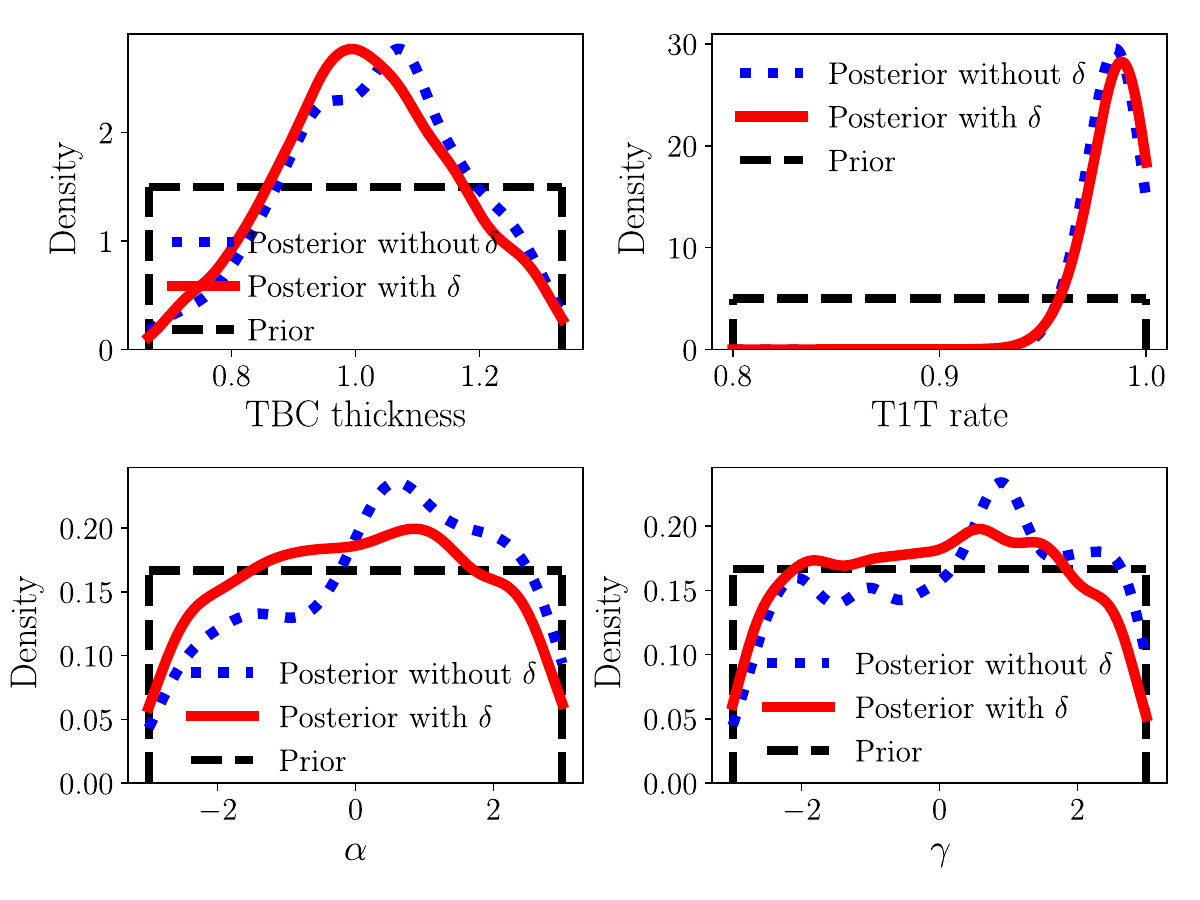}
  \caption{Posterior distributions of model parameters using elongation data at the blade tip (blade ID 90), using offline Bayesian calibration}
  \label{fig:modelparams_zdisp}
\end{figure}

\begin{figure}[t]
\centering
    \includegraphics[width=2.85in]{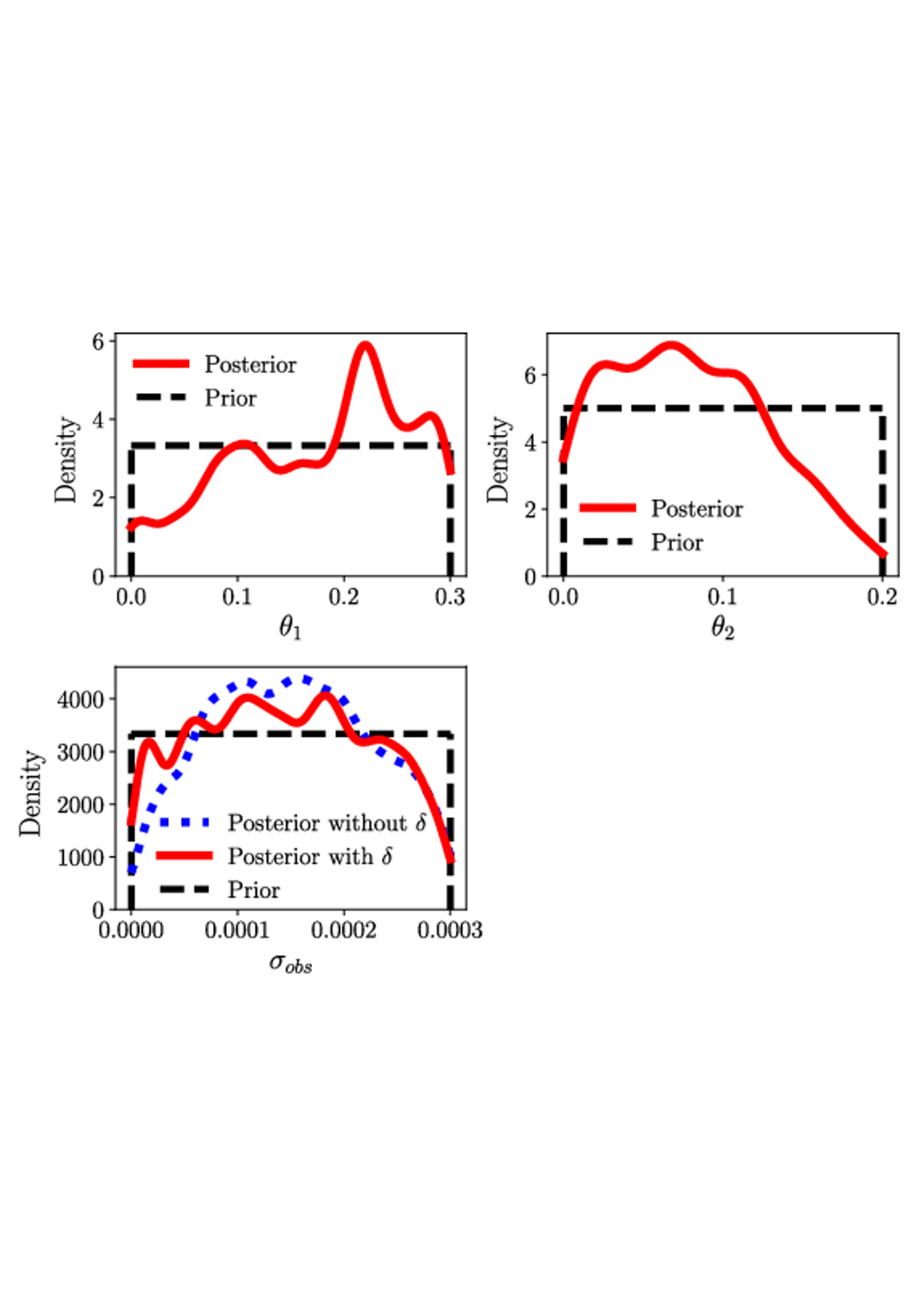}
  \caption{Posterior distributions of noise standard deviation and model discrepancy hyperparameters using elongation data at the blade tip (blade ID 90), using offline Bayesian calibration}
  \label{fig:hyperparams_zdisp}
\end{figure}

\begin{figure}[t]
\centering
    \includegraphics[width=2.85in]{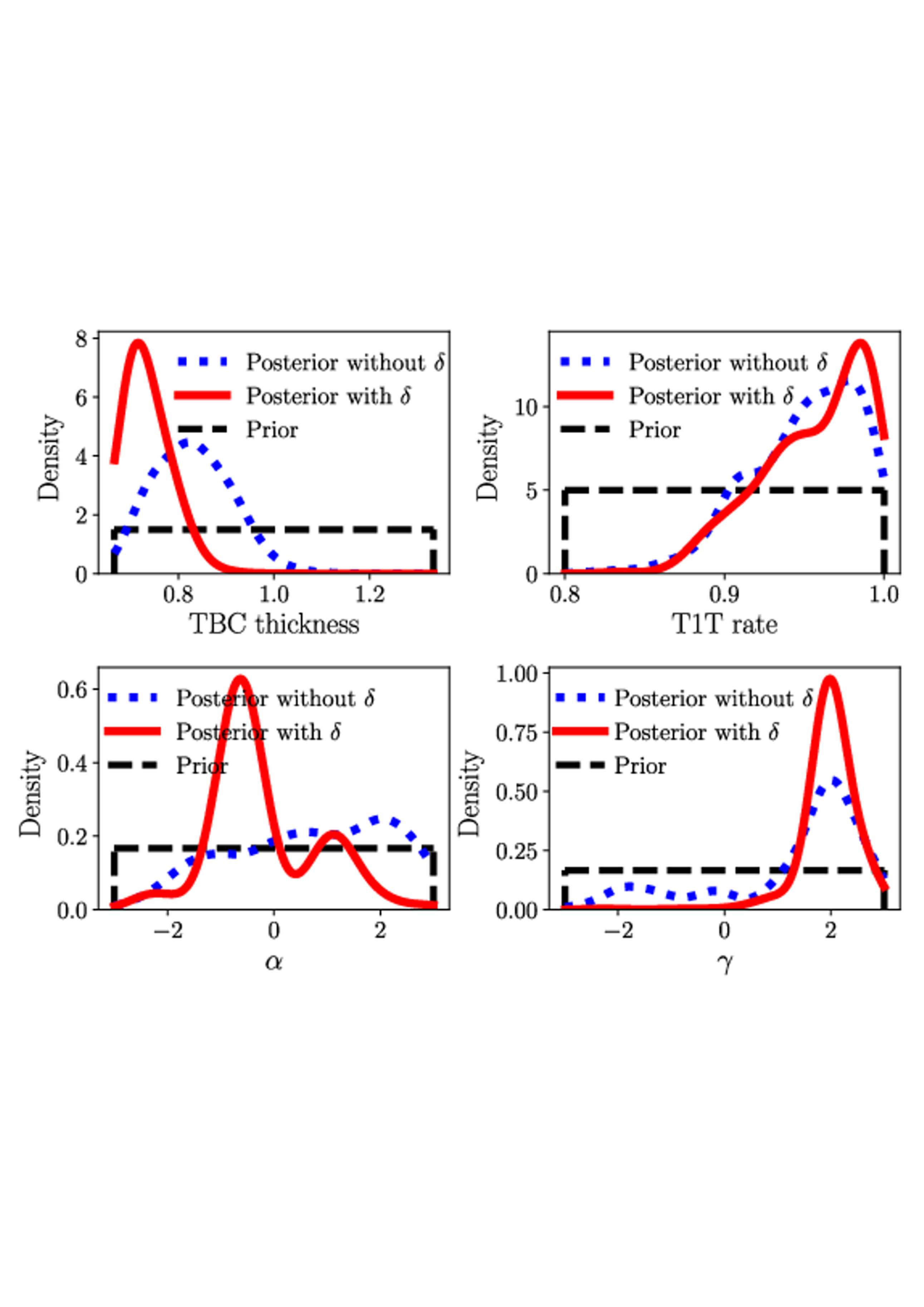}
  \caption{Posterior distributions of $\boldsymbol{\psi}$ using 3D deformation data of blade ID 01, using offline Bayesian calibration}
  \label{fig:modelparams_disp}
\end{figure}

\begin{figure}[t]
\centering
    \includegraphics[width=2.85in]{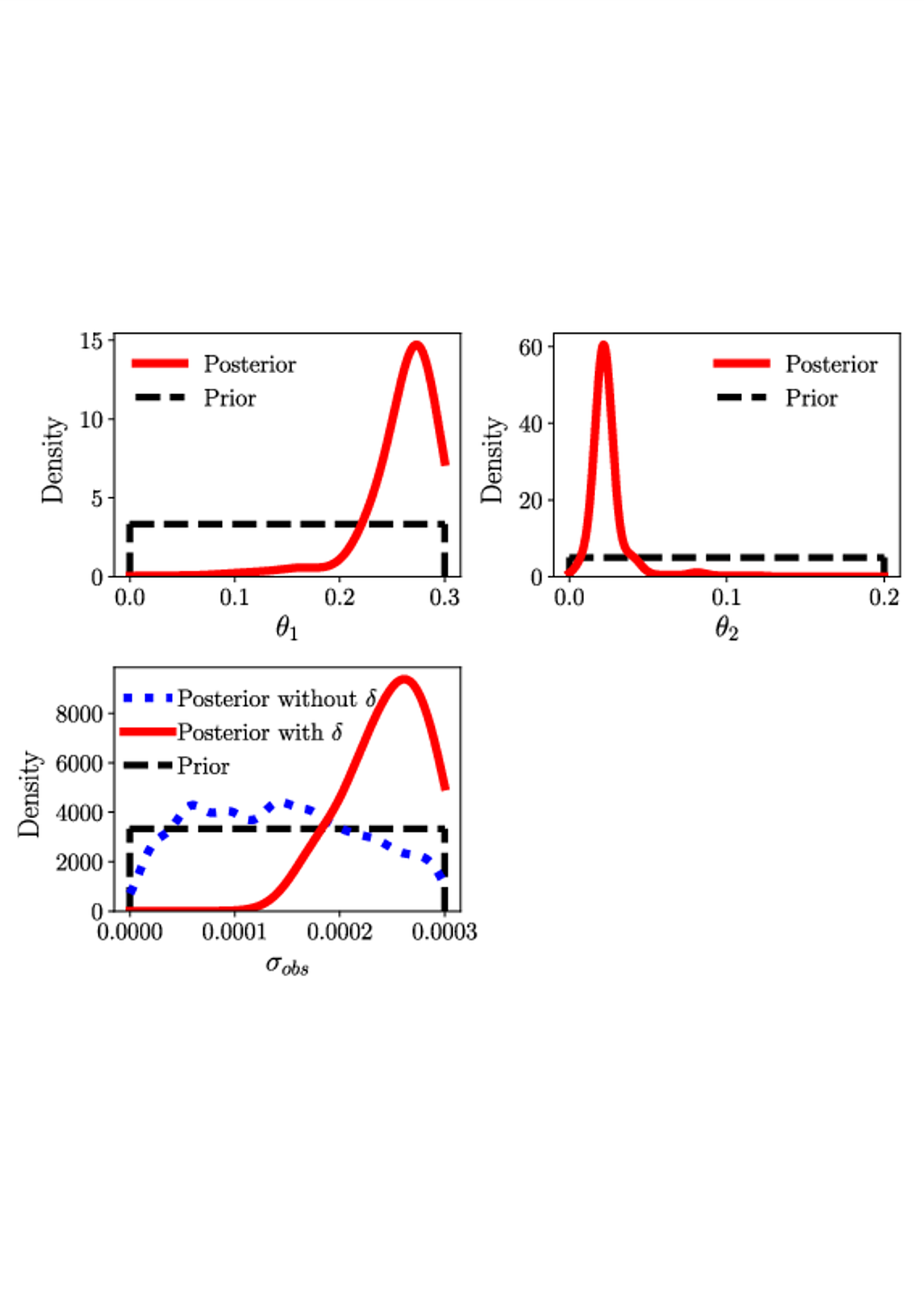}
  \caption{Posterior distributions of $\sigma_{obs}$ and $\boldsymbol{\theta}$ based on 3D deformation data of blade ID 01, using offline Bayesian calibration}
  \label{fig:hyperparams_disp}
\end{figure}

Two options of multi-level Bayesian calibration are pursued: offline and online. The resulting posterior distributions of parameters based on the offline Bayesian calibration corresponding to blade ID 90 (which showed medium creep damage) obtained by using the multi-blade measurement field data (i.e., elongation at the blade tip) are shown in Figs.~\ref{fig:modelparams_zdisp} and~\ref{fig:hyperparams_zdisp}. The calibration approach was performed on a blade with medium creep damage, i.e., blade ID 90. The posterior distributions of parameters corresponding to blade ID 01 obtained by using the 3D deformation data based on the offline Bayesian calibration are given in Figs.~\ref{fig:modelparams_disp} and~\ref{fig:hyperparams_disp}.

The true values of all parameters are unknown. The parameter ranges used for the priors are: TBC thickness [2/3,4/3], T1T rate [0.8,1], $\alpha$ and $\gamma$ [-3,3]. Generally, T1T is a known parameter in actual turbines. The proposed method utilizes (equivalent) T1T as a governing parameter of metal temperature of turbine blades, which has multiple uncertainty sources such as heat transfer coefficient in combustion gas and cooling air, etc. The results with and without a model discrepancy term (i.e., only considering measurement error $\epsilon_{obs}$) are plotted together. In the presence of insufficient amount of experimental data and non-informative prior knowledge about the uncertainty sources in the engineering system, it may be difficult to distinguish between the effects of the model parameters and model discrepancy as the number of parameters that need to be estimated becomes large; this problem is referred to as non-identifiability~\cite{ling2014selection,arendt2012quantification}. Hence, we assume a constant value for the standard deviation of model discrepancy (i.e., $\sigma_{\delta}(\boldsymbol{\psi},\boldsymbol{y},\boldsymbol{\theta})=\sigma_{\delta}=1\times10^{-5}$). We have tried several different model discrepancy functions. Based on the analysis of surrogate model predictions and measurement data the following model discrepancy formulation is chosen $\delta\sim N(\mu_{\delta}(\boldsymbol{\psi},\boldsymbol{y},\boldsymbol{\theta}),\sigma_{\delta}(\boldsymbol{\psi},\boldsymbol{y},\boldsymbol{\theta})$), with $\mu_{\delta}(\boldsymbol{\psi},\boldsymbol{y},\boldsymbol{\theta})=\boldsymbol{y}(\boldsymbol{\theta}_1-\boldsymbol{\theta}_2\boldsymbol{\psi})$. The bias is represented in terms of two parameters, $\boldsymbol{\theta}_1$ and $\boldsymbol{\theta}_2$. The bias term is dependent on the values of the actual model parameters and $\boldsymbol{\theta}_1$ and $\boldsymbol{\theta}_2$ help to exploit this dependency. Figs.~\ref{fig:modelparams_disp} and \ref{fig:hyperparams_disp} show that the inclusion of model discrepancy results in slightly sharper posterior distributions (indicating larger uncertainty reduction) for the calibration parameters as additional single-component data is used in re-calibration.

\begin{figure*}[t]
\centering
    \includegraphics[width=4.95in]{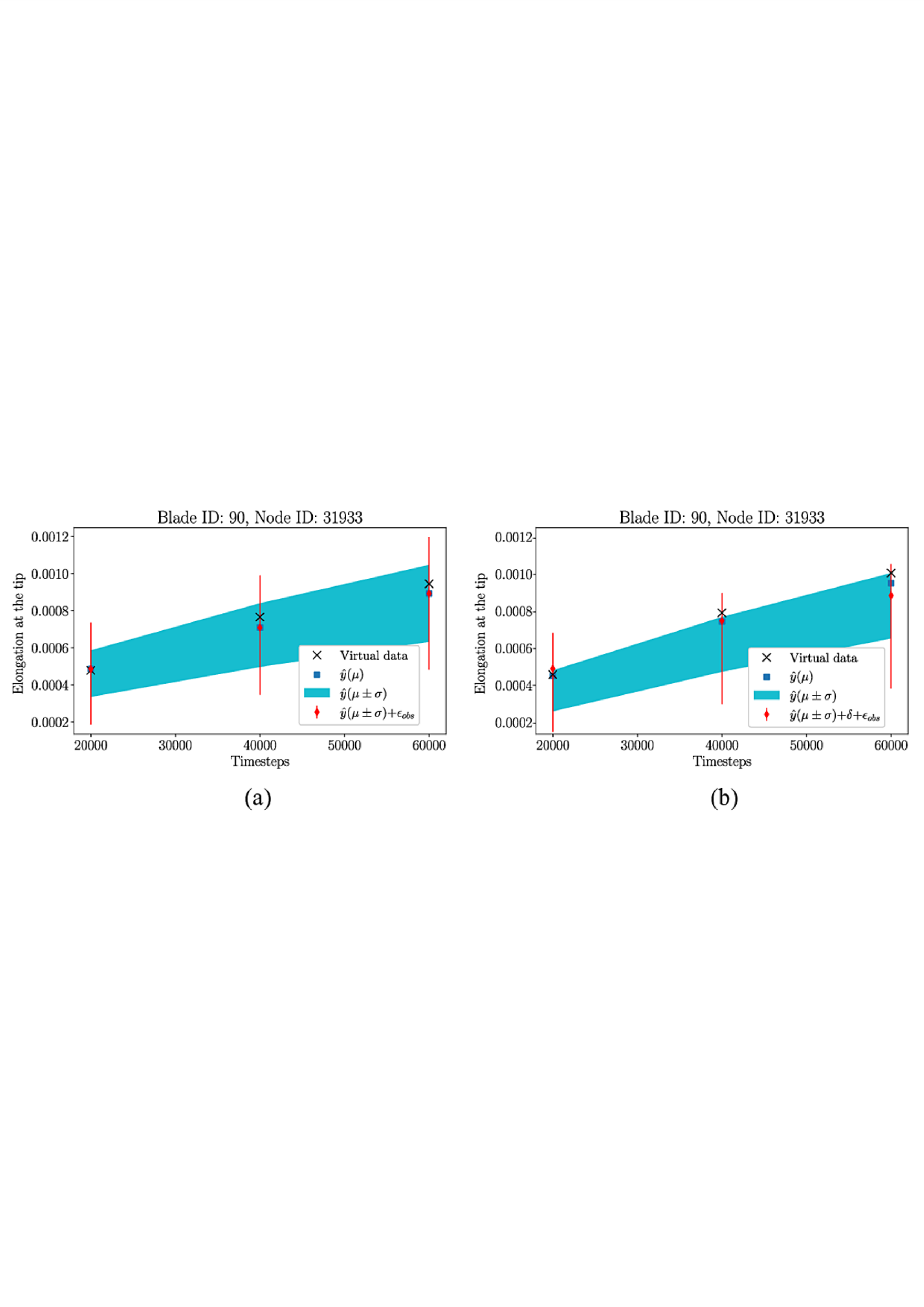}
  \caption{Prediction $\hat{y}$ at the mean and plus or minus one standard deviation of the calibration parameter estimates, for the elongation at the blade tip (blade ID 90), using offline Bayesian calibration: (a) without model discrepancy, (b) with model discrepancy}
  \label{fig:pred_zdisp}
\end{figure*}

\begin{figure*}[t]
\centering
    \includegraphics[width=4.95in]{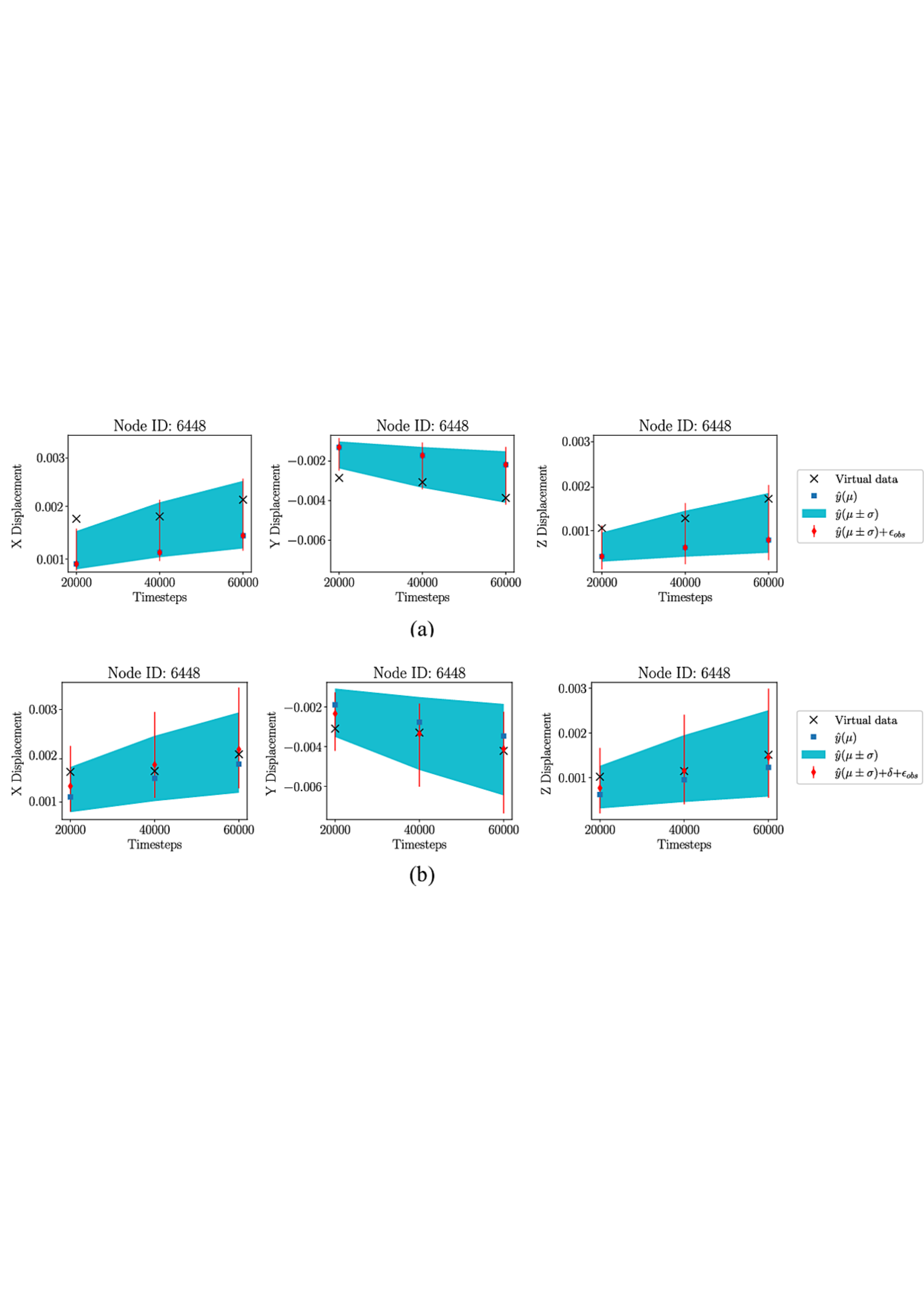}
  \caption{Prediction $\hat{y}$ at the mean and plus or minus one standard deviation of the calibrated parameters using the 3D deformation data of blade ID 01, using offline Bayesian calibration: (a) without model discrepancy, (b) with model discrepancy}
  \label{fig:pred_disp}
\end{figure*}

\begin{figure}[t]
\centering
    \includegraphics[width=2.85in]{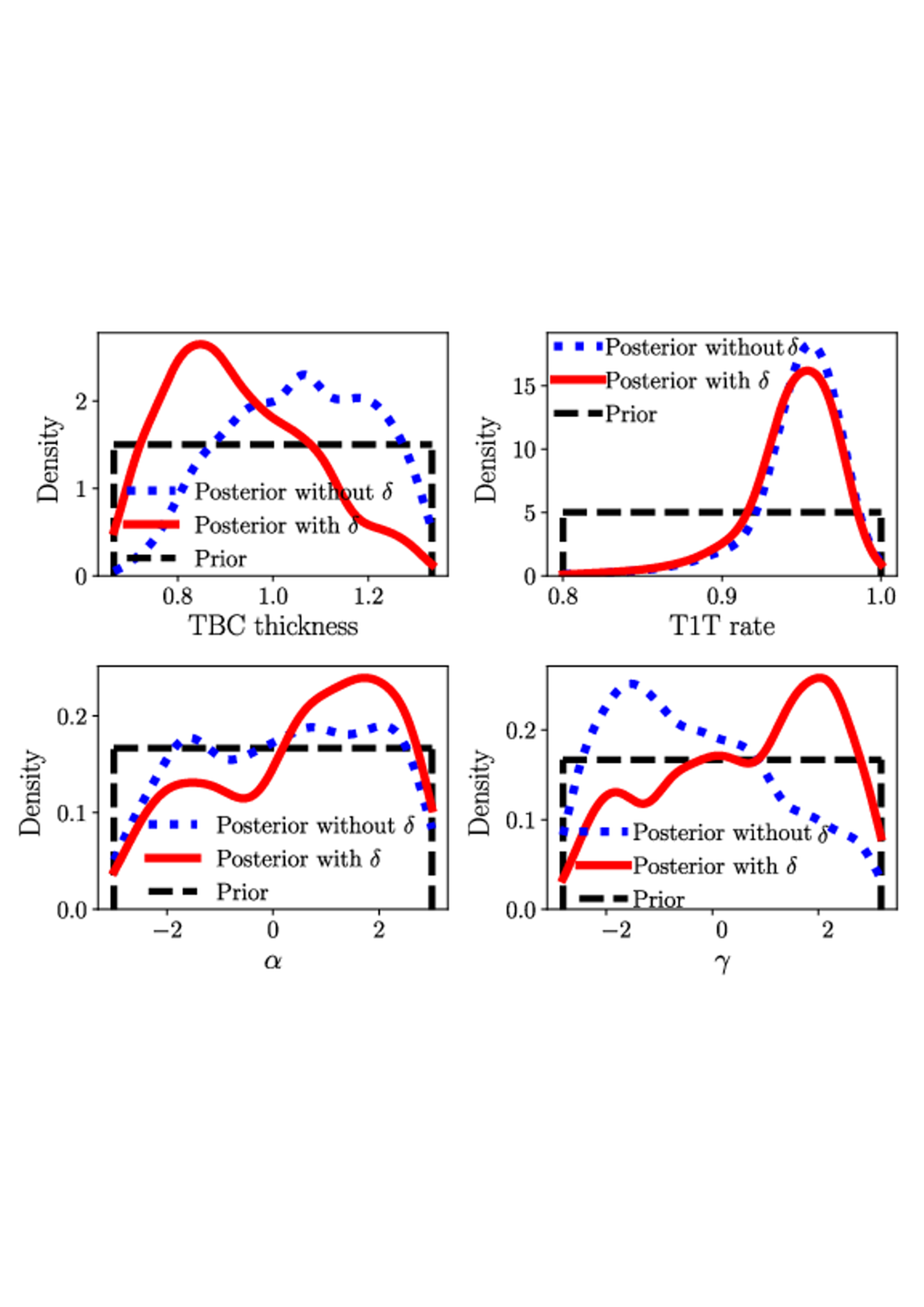}
  \caption{Posterior distributions of $\boldsymbol{\psi}$ using elongation data at the blade tip (blade ID 90), using online Bayesian calibration}
  \label{fig:modelparams_zdisp_insitu}
\end{figure}

\begin{figure}[t]
\centering
    \includegraphics[width=2.85in]{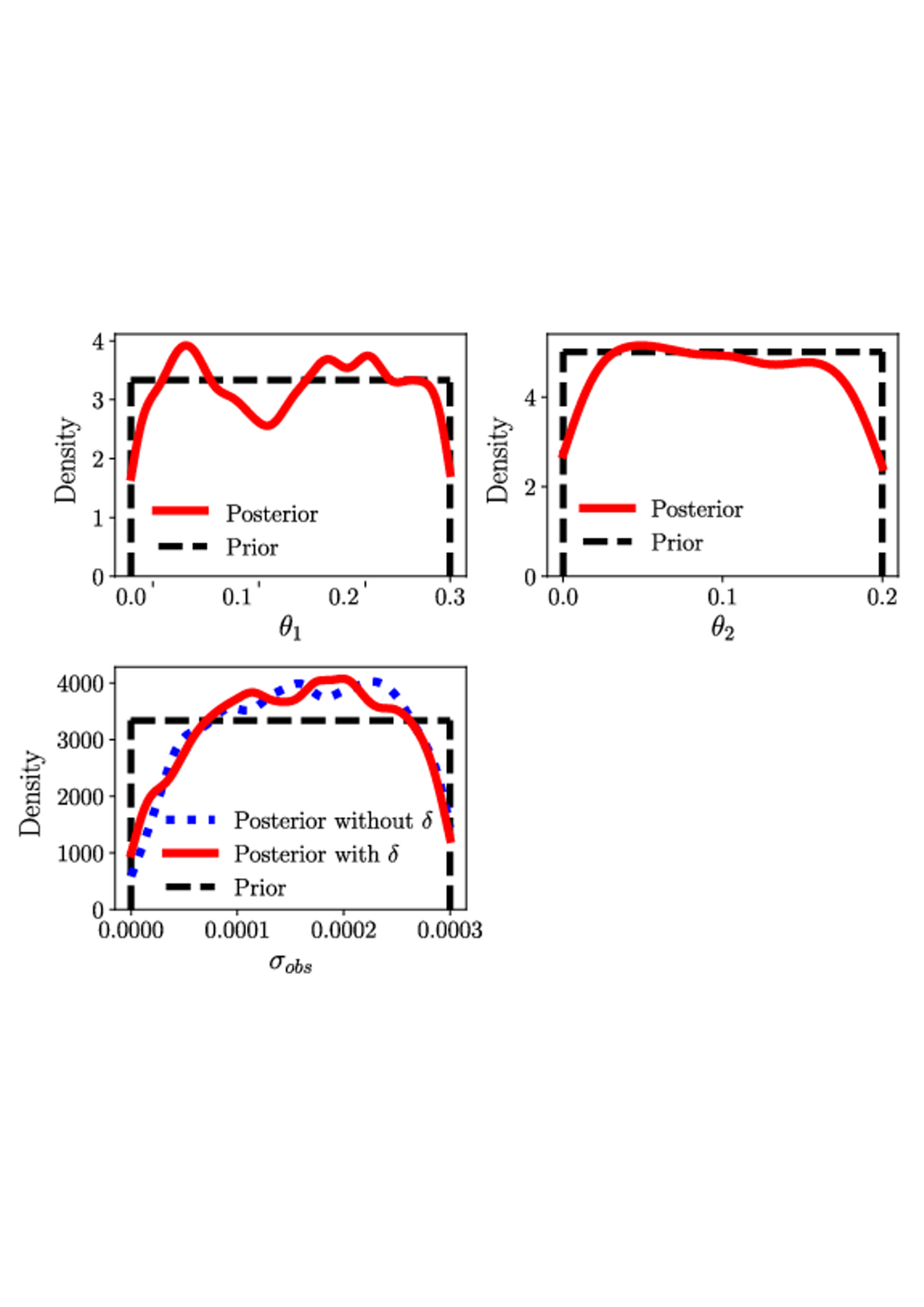}
  \caption{Posterior distributions of $\sigma_{obs}$ and $\boldsymbol{\theta}$ using elongation data at the blade tip (blade ID 90), using online Bayesian calibration}
  \label{fig:hyperparams_zdisp_insitu}
\end{figure}

\begin{figure}[t]
\centering
    \includegraphics[width=2.85in]{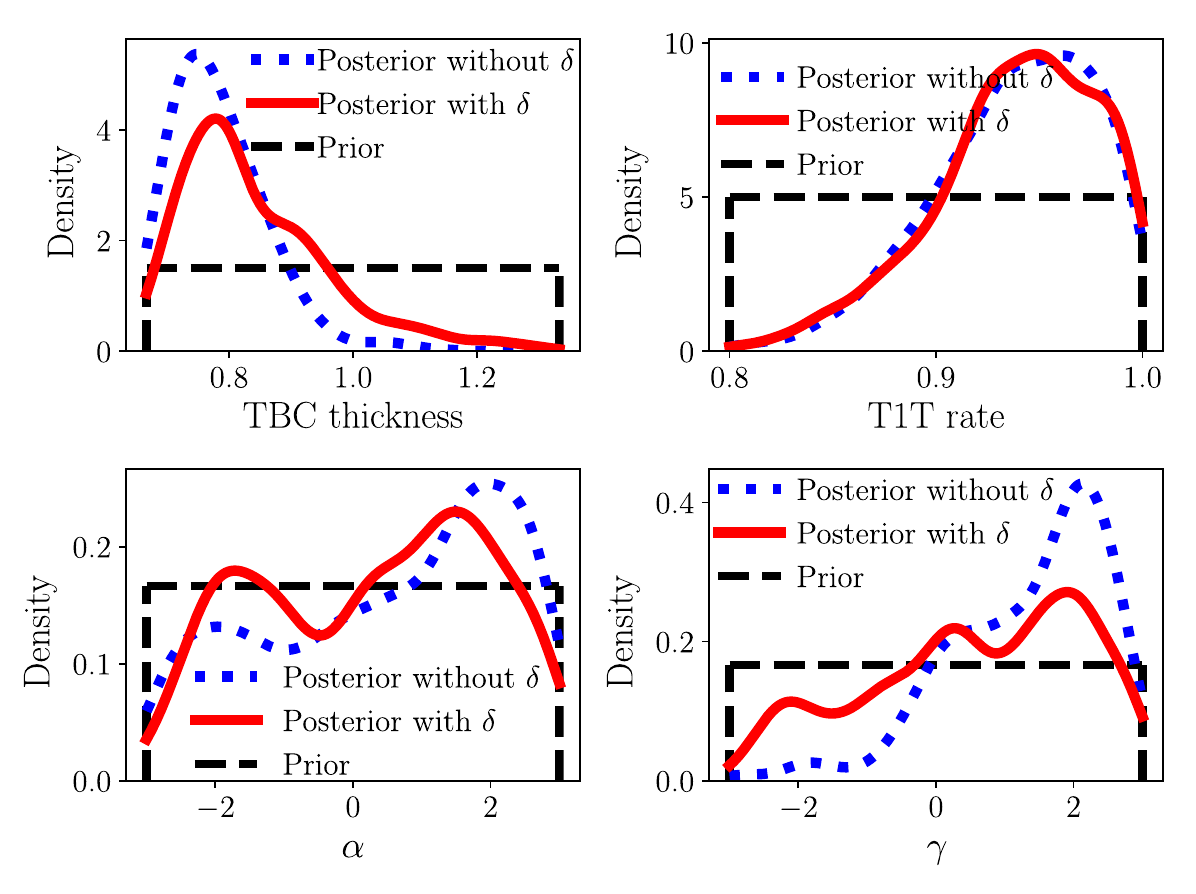}
  \caption{Posterior distributions of $\boldsymbol{\psi}$ using 3D deformation data of blade ID 01, using online Bayesian calibration}
  \label{fig:modelparams_disp_insitu}
\end{figure}

\begin{figure}[t]
\centering
    \includegraphics[width=2.85in]{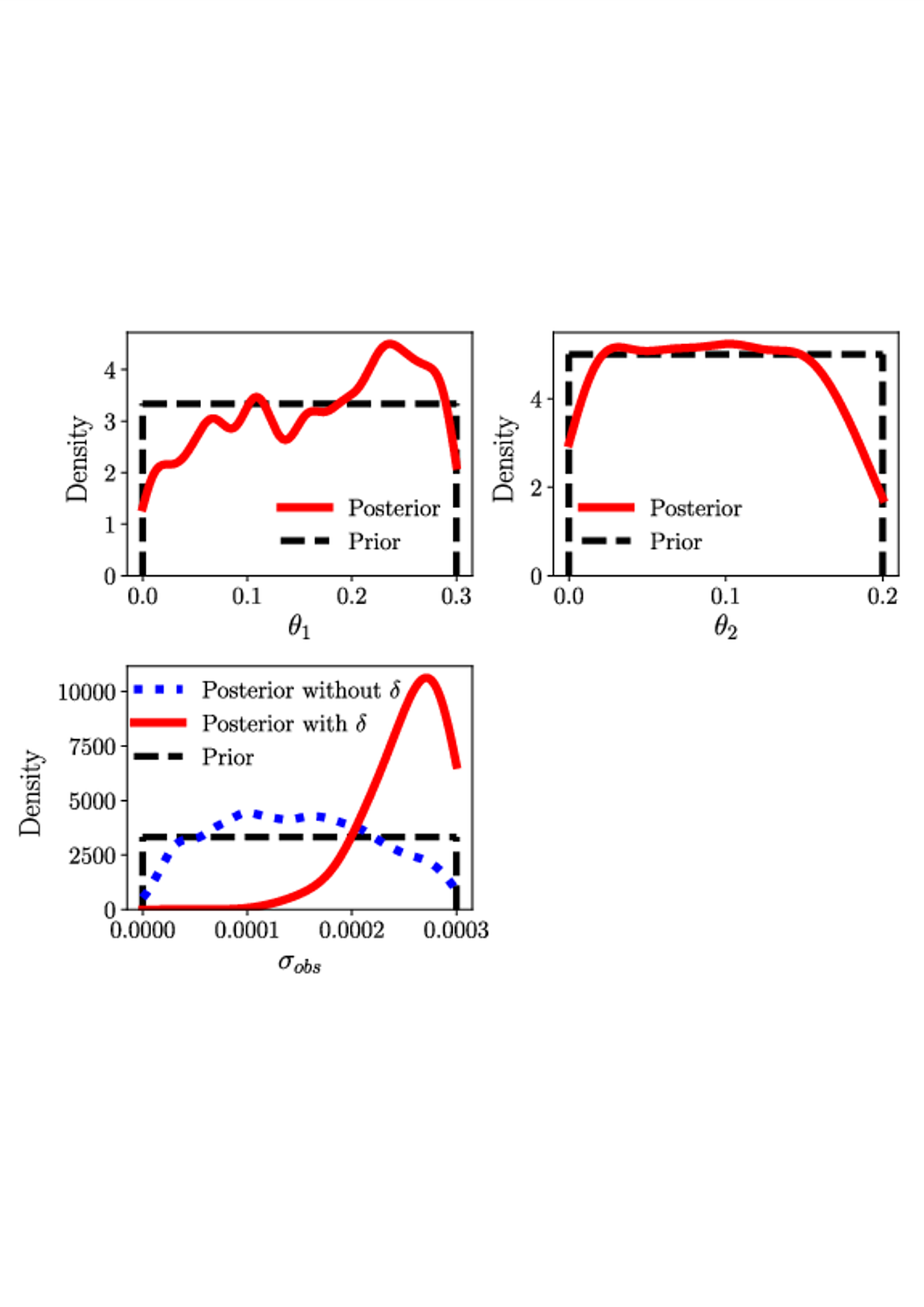}
  \caption{Posterior distributions of $\sigma_{obs}$ and $\boldsymbol{\theta}$ based on 3D deformation data of blade ID 01, using online Bayesian calibration}
  \label{fig:hyperparams_disp_insitu}
\end{figure}

\begin{figure*}[t]
\centering
    \includegraphics[width=4.in]{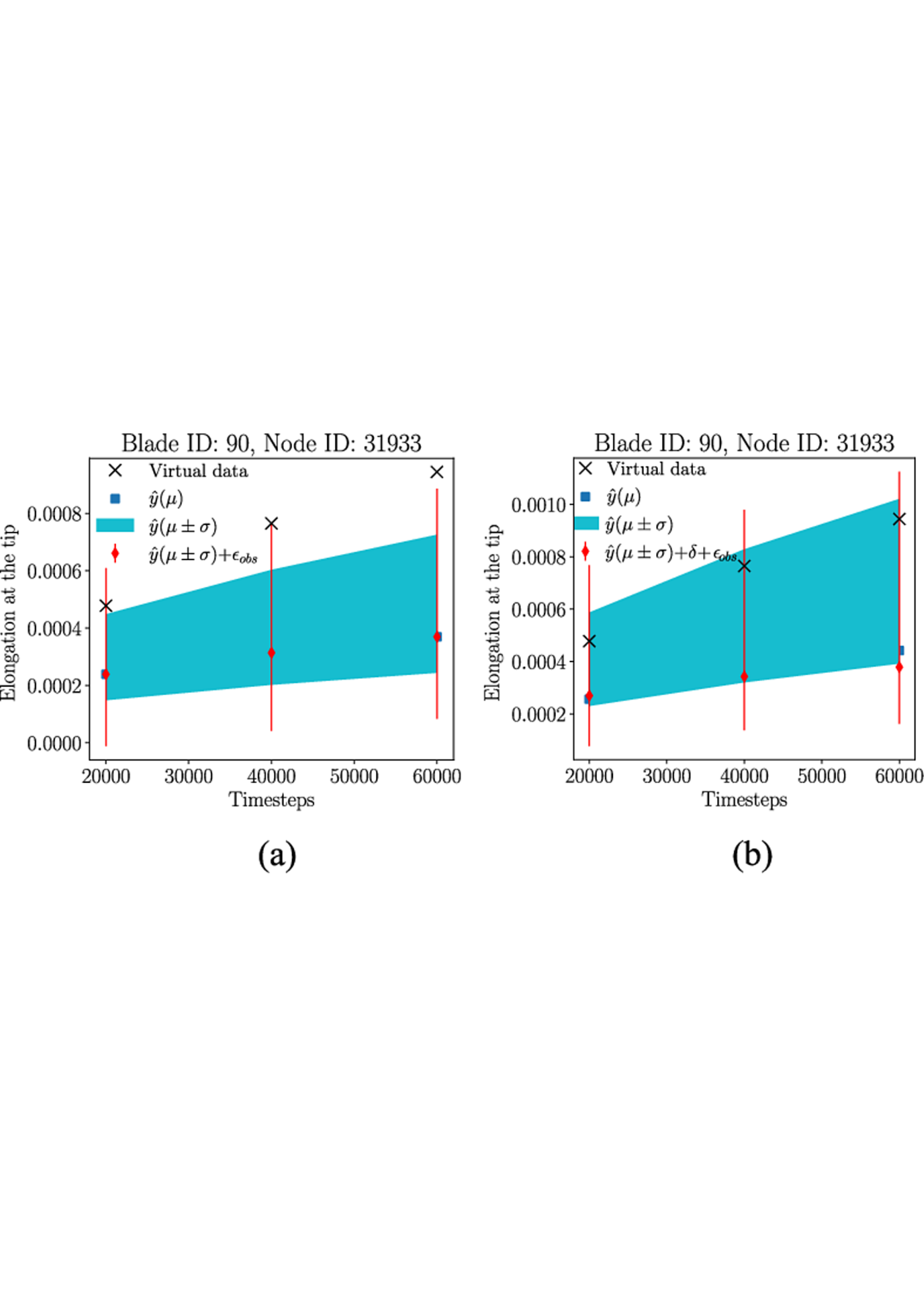}
  \caption{Prediction $\hat{y}$ at the mean and plus or minus one standard deviation of the calibration parameter estimates, for the elongation at the blade tip (blade ID 90), using online Bayesian calibration: (a) without model discrepancy, (b) with model discrepancy}
  \label{fig:pred_zdisp_insitu}
\end{figure*}

\begin{figure*}[t]
\centering
    \includegraphics[width=4.95in]{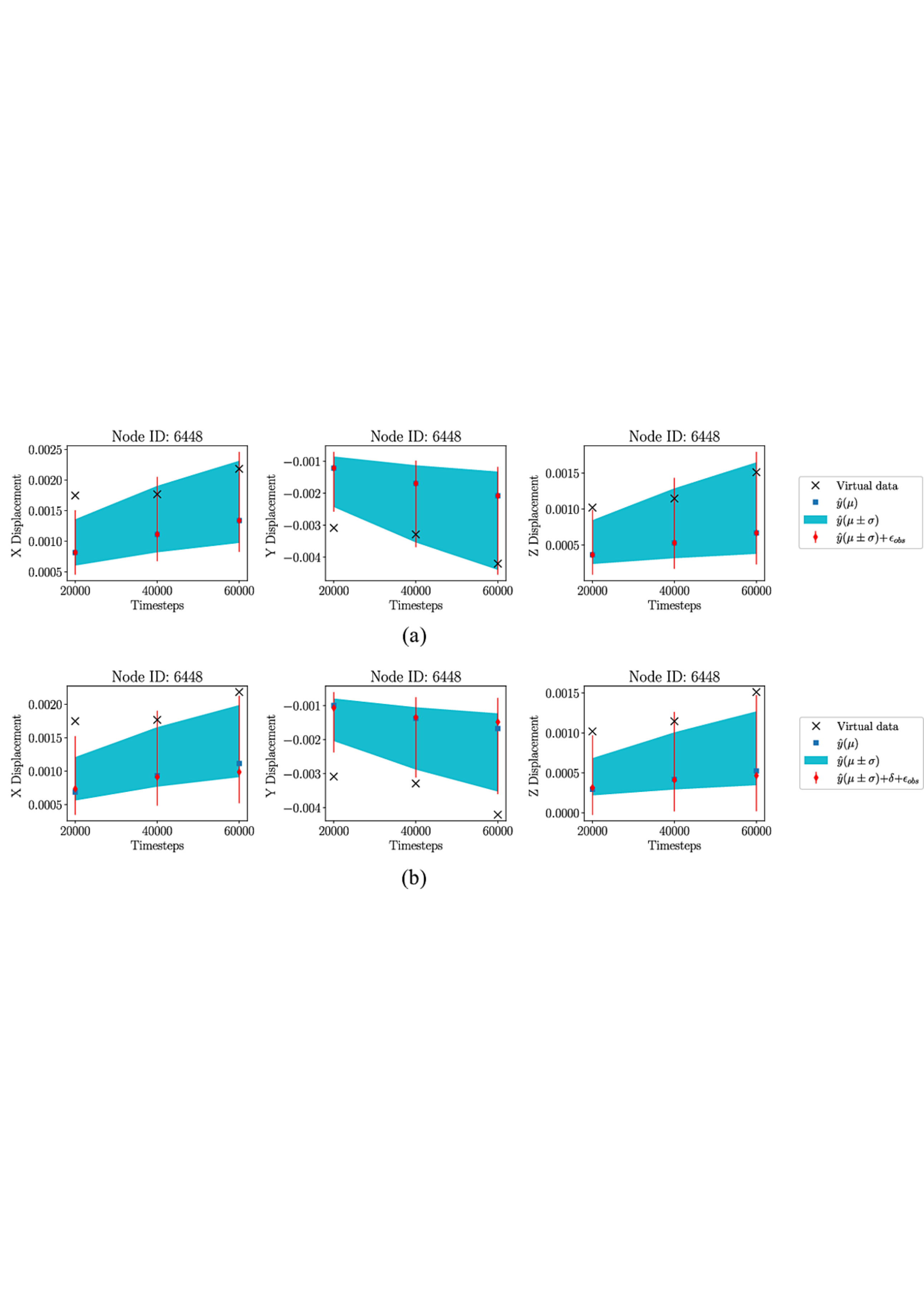}
  \caption{Prediction $\hat{y}$ at the mean and plus or minus one standard deviation of the calibrated parameters using the 3D deformation data of blade ID 01, using online Bayesian calibration: (a) without model discrepancy, (b) with model discrepancy}
  \label{fig:pred_disp_insitu}
\end{figure*}

The prediction from the calibrated model is validated with additional measurement data not used for calibration. The mean values and standard deviations of the calibrated parameters, which are obtained from the weights of particles representing samples of the joint distribution, are propagated through the surrogate model to obtain the mean plus/minus one standard deviation of the prediction (first-order approximation). The predictions of the calibrated surrogate model (using offline Bayesian calibration) are compared against the measurement data in Figs.~\ref{fig:pred_zdisp} and~\ref{fig:pred_disp} to validate the calibration results. The blue squares denote the mean surrogate model predictions. The light blue shaded region represents the surrogate model predictions using the mean plus or minus one standard deviation $(\mu\pm\sigma)$ of the calibration parameter estimates. Similarly, the red error bar indicates the surrogate model predictions using the mean plus or minus one standard deviation of the calibration parameter estimates plus the measurement error $(\mu\pm\sigma+\epsilon_{obs})$ in Figs.~\ref{fig:pred_zdisp}(a) and~\ref{fig:pred_disp}(a) and model discrepancy $(\mu\pm\sigma+\epsilon_{obs}+\delta)$ in Figs.~\ref{fig:pred_zdisp}(b) and~\ref{fig:pred_disp}(b). The surrogate model predictions for 3D deformation are improved with the inclusion of $\delta$ (Fig.~\ref{fig:pred_disp}(b)). The measurements at each time step are observed to be within the prediction bounds obtained using the posterior estimates of the calibration parameters, demonstrating the effectiveness of the proposed method.
The posterior distributions of parameters based on the online Bayesian calibration are shown in Figs.~\ref{fig:modelparams_zdisp_insitu}, \ref{fig:hyperparams_zdisp_insitu}, \ref{fig:modelparams_disp_insitu}, and \ref{fig:hyperparams_disp_insitu} for calibration with different types of measurement data. The calibrated surrogate model predictions are compared against the measurement data as shown in Figs.~\ref{fig:pred_zdisp_insitu} and~\ref{fig:pred_disp_insitu} to assess the performance of the proposed online strategy. In Fig.~{\ref{fig:pred_zdisp_insitu}}, the accuracy of including a model discrepancy term is actually better than without a model discrepancy. However the uncertainty increases with the inclusion of model discrepancy, since there are additional parameters to calibrate.

The offline strategy, in which data at all time steps are used at once to calibrate the model parameters, captures more information as expected. Thus, when we compare Figs. \ref{fig:modelparams_disp} and \ref{fig:hyperparams_disp} with Figs. \ref{fig:modelparams_disp_insitu} and \ref{fig:hyperparams_disp_insitu}, respectively, we observe that the posterior distributions obtained using the offline strategy are significantly sharper. However, online calibration can identify any changes in real-time, and is computationally less demanding. In addition, since the model discrepancy term introduces additional parameters to be calibrated and there is insufficient amount of data, some of the posterior distributions without $\delta$ are sharper than those with $\delta$ (see Fig.~{\ref{fig:modelparams_disp_insitu}}). As a result of this, the uncertainty was not adequately quantified for some of the online calibration cases such as Fig.~{\ref{fig:pred_disp_insitu}(b)}, where available data is limited. The computational times of the entire offline and online calibration (i.e., multi-level calibration using different types of measurement data) with 10,000 particles are approximately 48 and 27 minutes on a desktop with an Intel 8-core CPU with a 3.00 GHz base frequency and 8 GB memory, which includes surrogate model prediction for each particle at each measurement time step.

\section{Conclusion}\label{Sec:Conclusion}

This paper presented a multi-level Bayesian calibration framework for multi-component systems with time-dependent response. The calibration strategy fuses heterogeneous and asynchronous information, from different types of measurements on different system components at different time instants.

A surrogate model is first built to replace the expensive physics-based FEM for faster calibration; this surrogate model is built to predict a very high-dimensional output of multiple QoIs over a large number of spatial locations and time instants. Next, a Hierarchical Bayesian Network (HBN) representation is developed to fuse different types of measurement data from different components at different time steps, thus help in to incorporate all the available information into the calibration strategy.

The multi-level hierarchy can be in terms of models or data types. The numerical example consists of two levels in terms of both models and data, and the proposed approach is able to accommodate both types of hierarchy. The multi-component data (elongation) is used prior to the single component data (deformation) because we had limited amount of single-component data to perform an effective calibration. The effect of multi-level calibration order is not significant since there are only three local parameters that are most directly influenced by the single-component (blade ID 1). If we had adequate amount of single-component data, then calibrating each local parameter with their corresponding single-component data before calibrating with multi-component data could have been more efficient (less particles needed). The overall calibration approach was shown in Fig. \ref{fig:steps_pf_multilevel}(b), and can be summarized as follows:
\begin{enumerate}
    \item Both local and shared parameters are calibrated with multi-component data using the iterative strategy described in Section \ref{sec:off_on}.
    \item Next, the local parameters corresponding to the specific components where additional data is available are re-calibrated using the single-component data.
\end{enumerate}
Both offline and online calibration strategies are investigated. The offline strategy, where the multi-component calibration is performed with data from all time steps (unlike the online strategy), converges to sharper posterior distributions. 

The proposed online approach could support building a \emph{digital twin} that contains all the information about the engineering system and is periodically updated as new information becomes available. The online calibration strategy developed above is particularly useful in this regard, by tracking the evolution of time-varying parameters. The digital twin can be used to (1) assess the current condition and capabilities of the system and its components, (2) predict the future condition and capabilities of the system and its components, and (3) provide information for decision-making related to system health management (inspection, maintenance, and repair) as well as operational control. Future work could consider additional measurements at several spatio-temporal locations to obtain sharper posterior estimates by fusing more information from multiple levels. Further investigation is needed regarding resource allocation and scheduling for data collection in the context of multi-level calibration and its connection to the calibration strategy (i.e., online or offline).

\section*{Acknowledgment}
This work is funded by Mitsubishi Heavy Industries in Nagasaki and Takasago, Japan. The support is gratefully acknowledged.

\bibliographystyle{unsrt}  
\bibliography{references}

@article{halko2011finding,
  title={Finding structure with randomness: Probabilistic algorithms for constructing approximate matrix decompositions},
  author={Halko, Nathan and Martinsson, Per-Gunnar and Tropp, Joel A},
  journal={SIAM review},
  volume={53},
  number={2},
  pages={217--288},
  year={2011},
  doi={10.1137/090771806},
  publisher={SIAM}
}

@article{ling2014selection,
  title={Selection of model discrepancy priors in Bayesian calibration},
  author={Ling, You and Mullins, Joshua and Mahadevan, Sankaran},
  journal={Journal of Computational Physics},
  volume={276},
  pages={665--680},
  year={2014},
  doi={10.1016/j.jcp.2014.08.005},
  publisher={Elsevier}
}

@article{morris1995exploratory,
  title={Exploratory designs for computational experiments},
  author={Morris, Max D and Mitchell, Toby J},
  journal={Journal of Statistical Planning and Inference},
  volume={43},
  number={3},
  pages={381--402},
  year={1995},
  doi={10.1016/0378-3758(94)00035-T},
  publisher={Elsevier}
}

@article{dempster1977maximum,
  title={Maximum likelihood from incomplete data via the EM algorithm},
  author={Dempster, Arthur P and Laird, Nan M and Rubin, Donald B},
  journal={Journal of the Royal Statistical Society: Series B (Methodological)},
  volume={39},
  number={1},
  pages={1--22},
  year={1977},
  doi={10.1111/j.2517-6161.1977.tb01600.x},
  publisher={Wiley Online Library}
}

@article{hastings1970monte,
    author = {Hastings, W. K.},
    title = "{Monte Carlo sampling methods using Markov chains and their applications}",
    journal = {Biometrika},
    volume = {57},
    number = {1},
    pages = {97-109},
    year = {1970},
    month = {04},
    issn = {0006-3444},
    doi = {10.1093/biomet/57.1.97},
}

@article{decarlo2016segmented,
  title={Segmented Bayesian calibration of multidisciplinary models},
  author={DeCarlo, Erin C and Smarslok, Benjamin P and Mahadevan, Sankaran},
  journal={AIAA Journal},
  volume={54},
  number={12},
  pages={3727--3741},
  year={2016},
  doi={10.2514/1.J054960},
  publisher={American Institute of Aeronautics and Astronautics}
}

@article{vanderhorn2018bayesian,
  title={Bayesian model updating with summarized statistical and reliability data},
  author={VanDerHorn, Eric and Mahadevan, Sankaran},
  journal={Reliability Engineering \& System Safety},
  volume={172},
  pages={12--24},
  year={2018},
  doi={10.1016/j.ress.2017.11.023},
  publisher={Elsevier}
}

@book{pearl1988probabilistic,
author = {Pearl, Judea},
title = {Probabilistic Reasoning in Intelligent Systems: Networks of Plausible Inference},
year = {1988},
isbn = {1558604790},
publisher = {Morgan Kaufmann Publishers Inc.},
address = {San Francisco, CA, USA}
}

@article{karve2020digital,
  title={Digital twin approach for damage-tolerant mission planning under uncertainty},
  author={Karve, Pranav M and Guo, Yulin and Kapusuzoglu, Berkcan and Mahadevan, Sankaran and Haile, Mulugeta A},
  journal={Engineering Fracture Mechanics},
  volume={225},
  pages={106766},
  year={2020},
  doi={10.1016/j.engfracmech.2019.106766},
  publisher={Elsevier}
}

@article{kennedy2001bayesian,
  title={Bayesian calibration of computer models},
  author={Kennedy, Marc C and O'Hagan, Anthony},
  journal={Journal of the Royal Statistical Society: Series B (Statistical Methodology)},
  volume={63},
  number={3},
  pages={425--464},
  year={2001},
  doi={10.1111/1467-9868.00294},
  publisher={Wiley Online Library}
}

@inproceedings{nannapaneni2016manufacturing,
    author = {Nannapaneni, Saideep and Mahadevan, Sankaran},
    title = {Manufacturing Process Evaluation Under Uncertainty: A Hierarchical Bayesian Network Approach},
    volume = {Volume 1B: 36th Computers and Information in Engineering Conference},
    booktitle = {IDETC-CIE},
    pages={V01BT02A026},
    year = {2016},
    month = {08},
    organization={ASME}
}

@article{xiu2002wiener,
  title={The Wiener--Askey polynomial chaos for stochastic differential equations},
  author={Xiu, Dongbin and Karniadakis, George Em},
  journal={SIAM Journal on Scientific Computing},
  volume={24},
  number={2},
  pages={619--644},
  year={2002},
  doi={10.1137/S1064827501387826},
  publisher={SIAM}
}

@Inbook{Rasmussen2004,
author="Rasmussen, Carl Edward",
title="Gaussian Processes in Machine Learning",
bookTitle="Advanced Lectures on Machine Learning: ML Summer Schools 2003, Canberra, Australia, February 2 - 14, 2003, T{\"u}bingen, Germany, August 4 - 16, 2003, Revised Lectures",
year="2004",
publisher="Springer Berlin Heidelberg",
address="Berlin, Heidelberg",
pages={63--71},
doi="10.1007/978-3-540-28650-9_4"
}

@article{arendt2012quantification,
    author = {Arendt, Paul D. and Apley, Daniel W. and Chen, Wei},
    title = "{Quantification of Model Uncertainty: Calibration, Model Discrepancy, and Identifiability}",
    journal = {Journal of Mechanical Design},
    volume = {134},
    number = {10},
    year = {2012},
    month = {09},
    issn = {1050-0472},
    doi = {10.1115/1.4007390},
}

@article{li2016role,
  title={Role of calibration, validation, and relevance in multi-level uncertainty integration},
  author={Li, Chenzhao and Mahadevan, Sankaran},
  journal={Reliability Engineering \& System Safety},
  volume={148},
  pages={32--43},
  year={2016},
  doi={10.1016/j.ress.2015.11.013},
  publisher={Elsevier}
}

@article{mullins2016bayesian,
  title={Bayesian uncertainty integration for model calibration, validation, and prediction},
  author={Mullins, Joshua and Mahadevan, Sankaran},
  journal={Journal of Verification, Validation and Uncertainty Quantification},
  volume={1},
  number={1},
  year={2016},
  doi={10.1115/1.4032371},
  publisher={ASME}
}

@article{sankararaman2015integration,
  title={Integration of model verification, validation, and calibration for uncertainty quantification in engineering systems},
  author={Sankararaman, Shankar and Mahadevan, Sankaran},
  journal={Reliability Engineering \& System Safety},
  volume={138},
  pages={194--209},
  year={2015},
  doi={10.1016/j.ress.2015.01.023},
  publisher={Elsevier}
}

@article{mahadevan2001bayesian,
  title={Bayesian networks for system reliability reassessment},
  author={Mahadevan, Sankaran and Zhang, Ruoxue and Smith, Natasha},
  journal={Structural Safety},
  volume={23},
  number={3},
  pages={231--251},
  year={2001},
  publisher={Elsevier}
}

@article{fearnhead2004particle,
  title={Particle filters for mixture models with an unknown number of components},
  author={Fearnhead, Paul},
  journal={Statistics and Computing},
  volume={14},
  number={1},
  pages={11--21},
  year={2004},
  doi={10.1023/B:STCO.0000009418.04621.cd},
  publisher={Springer}
}

@inproceedings{NIPS2017_6449f44a,
 author = {Ke, Guolin and Meng, Qi and Finley, Thomas and Wang, Taifeng and Chen, Wei and Ma, Weidong and Ye, Qiwei and Liu, Tie-Yan},
 booktitle = {Advances in Neural Information Processing Systems},
 editor = {I. Guyon and U. V. Luxburg and S. Bengio and H. Wallach and R. Fergus and S. Vishwanathan and R. Garnett},
 pages = {3149–3157},
 publisher = {Curran Associates, Inc.},
 title = {LightGBM: A Highly Efficient Gradient Boosting Decision Tree},
 volume = {30},
 year = {2017}
}

@inproceedings{xgboost,
author = {Chen, Tianqi and Guestrin, Carlos},
title = {XGBoost: A Scalable Tree Boosting System},
booktitle={Proceedings of the 22nd ACM SIGKDD International Conference on Knowledge Discovery and Data Mining},
year = {2016},
isbn = {9781450342322},
publisher = {Association for Computing Machinery},
address = {New York, NY, USA},
doi = {10.1145/2939672.2939785},
pages = {785–794},
numpages = {10},
location = {San Francisco, California, USA},
series = {KDD '16}
}

@inproceedings{dorogush2018catboost,
 author = {Prokhorenkova, Liudmila and Gusev, Gleb and Vorobev, Aleksandr and Dorogush, Anna Veronika and Gulin, Andrey},
 booktitle = {Advances in Neural Information Processing Systems},
 editor = {S. Bengio and H. Wallach and H. Larochelle and K. Grauman and N. Cesa-Bianchi and R. Garnett},
 pages = {},
 publisher = {Curran Associates, Inc.},
 title = {CatBoost: unbiased boosting with categorical features},
 volume = {31},
 year = {2018}
}

@article{geurts2006extremely,
  title={Extremely randomized trees},
  author={Geurts, Pierre and Ernst, Damien and Wehenkel, Louis},
  journal={Machine Learning},
  volume={63},
  number={1},
  pages={3--42},
  doi={10.1007/s10994-006-6226-1},
  year={2006},
  publisher={Springer}
}

@article{hastie2009elements,
  title={The elements of statistical learning: data mining, inference and prediction},
  author={Franklin, James},
  journal={The Mathematical Intelligencer},
  volume={27},
  number={2},
  pages={83--85},
  year={2005},
  publisher={Springer}
}

@inproceedings{kapusuzoglu2022dimension,
  title={Dimension Reduction for Efficient Surrogate Modeling in High-Dimensional Applications},
  author={Kapusuzoglu, Berkcan and Guo, Yulin and Mahadevan, Sankaran and Matsumoto, Shunsaku and Yoshitomo, Miyagi and Taba, Shunsuke and Watanabe, Daigo},
  booktitle={AIAA SCITECH 2022 Forum},
  pages={1440},
  doi={10.2514/6.2022-1440},
  year={2022}
}

@article{Moral2006,
author = {Del Moral, Pierre and Doucet, Arnaud and Jasra, Ajay},
title = {Sequential Monte Carlo samplers},
journal = {Journal of the Royal Statistical Society: Series B (Statistical Methodology)},
volume = {68},
number = {3},
pages = {411-436},
doi = {https://doi.org/10.1111/j.1467-9868.2006.00553.x},
year = {2006}
}

@article{rebba2006model,
  title={Model predictive capability assessment under uncertainty},
  author={Rebba, Ramesh and Mahadevan, Sankaran},
  journal={AIAA journal},
  volume={44},
  number={10},
  pages={2376--2384},
  doi={10.2514/1.19103},
  year={2006}
}

@article{song2020accounting,
  title={Accounting for modeling errors and inherent structural variability through a hierarchical bayesian model updating approach: an overview},
  author={Song, Mingming and Behmanesh, Iman and Moaveni, Babak and Papadimitriou, Costas},
  journal={Sensors},
  volume={20},
  number={14},
  pages={3874},
  year={2020},
  publisher={MDPI}
}

@article{sedehi2019probabilistic,
  title={Probabilistic hierarchical Bayesian framework for time-domain model updating and robust predictions},
  author={Sedehi, Omid and Papadimitriou, Costas and Katafygiotis, Lambros S},
  journal={Mechanical Systems and Signal Processing},
  volume={123},
  pages={648--673},
  year={2019},
  doi={10.1016/j.ymssp.2018.09.041},
  publisher={Elsevier}
}

@article{SEDEHI2020106663,
title = {Hierarchical Bayesian operational modal analysis: Theory and computations},
journal = {Mechanical Systems and Signal Processing},
volume = {140},
pages = {106663},
year = {2020},
issn = {0888-3270},
doi = {10.1016/j.ymssp.2020.106663},
author = {Omid Sedehi and Lambros S. Katafygiotis and Costas Papadimitriou},
}

@article{behmanesh2015hierarchical,
  title={Hierarchical Bayesian model updating for structural identification},
  author={Behmanesh, Iman and Moaveni, Babak and Lombaert, Geert and Papadimitriou, Costas},
  journal={Mechanical Systems and Signal Processing},
  volume={64},
  pages={360--376},
  year={2015},
  doi={10.1016/j.ymssp.2015.03.026},
  publisher={Elsevier}
}

@article{nagel2015bayesian,
  title={Bayesian multilevel model calibration for inverse problems under uncertainty with perfect data},
  author={Nagel, Joseph B and Sudret, Bruno},
  journal={Journal of Aerospace Information Systems},
  volume={12},
  number={1},
  pages={97--113},
  year={2015},
  doi={10.2514/1.I010264},
  publisher={American Institute of Aeronautics and Astronautics}
}

@article{nagel2016unified,
  title={A unified framework for multilevel uncertainty quantification in Bayesian inverse problems},
  author={Nagel, Joseph B and Sudret, Bruno},
  journal={Probabilistic Engineering Mechanics},
  volume={43},
  pages={68--84},
  year={2016},
  doi={10.1016/j.probengmech.2015.09.007},
  publisher={Elsevier}
}

@article{jia2022hierarchical,
  title={Hierarchical Bayesian modeling framework for model updating and robust predictions in structural dynamics using modal features},
  author={Jia, Xinyu and Sedehi, Omid and Papadimitriou, Costas and Katafygiotis, Lambros S and Moaveni, Babak},
  journal={Mechanical Systems and Signal Processing},
  volume={170},
  pages={108784},
  year={2022},
  doi={10.1016/j.ymssp.2021.108784},
  publisher={Elsevier}
}

@article{decarlo2018quantifying,
  title={Quantifying model discrepancy in time-dependent, coupled analyses},
  author={DeCarlo, Erin C and Smarslok, Benjamin P and Mahadevan, Sankaran},
  journal={AIAA Journal},
  volume={56},
  number={6},
  pages={2403--2411},
  year={2018},
  doi={10.2514/1.J056719},
  publisher={American Institute of Aeronautics and Astronautics}
}

@article{viana2021survey,
  title={A survey of Bayesian calibration and physics-informed neural networks in scientific modeling},
  author={Viana, Felipe AC and Subramaniyan, Arun K},
  journal={Archives of Computational Methods in Engineering},
  volume={28},
  number={5},
  pages={3801--3830},
  year={2021},
  doi={10.1007/s11831-021-09539-0},
  publisher={Springer}
}

\end{document}